\documentclass[12pt]{article}
\usepackage{amsmath}
\usepackage{amssymb}
\usepackage{graphics}
\usepackage[all]{xy}

\makeatletter\@addtoreset{equation}{section}
\makeatother
\allowdisplaybreaks[1]
\begin{document}
\begin{titlepage}
\begin{flushright}
TIT/HEP-630\\
July 2013
\end{flushright}
\vspace{0.5cm}
\begin{center}
{\Large \bf
Twisted $\mathcal{N} = 4$ Super Yang-Mills Theory \\ 
\vspace{0.1cm}
in $\Omega$-background
}
\lineskip .75em
\vskip0.5cm
{\large Katsushi Ito${}^{1}$, Hiroaki Nakajima${}^{2}$ and Shin
Sasaki${}^{3}$ }
\vskip 2.5em
${}^{1}$ {\normalsize\it Department of Physics,\\
Tokyo Institute of Technology\\
Tokyo, 152-8551, Japan} \vskip 1.0em
${}^{2}$ {\normalsize\it Department of Physics 
and Center for Theoretical Sciences,\\
National Taiwan University\\
Taipei, 10617, Taiwan, R.O.C.} \vskip 1.0em
${}^{3}$ {\normalsize\it Department of Physics,\\
Kitasato University\\
Sagamihara, 252-0373, Japan}
\vskip 3.0em
\end{center}
\begin{abstract}
We study the twisted $\mathcal{N} = 4$ super Yang-Mills 
theories in the $\Omega$-background with the constant R-symmetry Wilson
 line gauge field. 
Based on the classification of topological twists of $\mathcal{N} = 4$ supersymmetry
(the half, the Vafa-Witten and the Marcus twists), 
we construct the deformed off-shell supersymmetry 
associated with the scalar supercharges for these twists.
We find that the $\Omega$-deformed action 
is written in the exact form with respect to the 
scalar supercharges as in the undeformed case.
\end{abstract}
\end{titlepage}

\baselineskip=0.7cm
\section{Introduction}
The $\Omega$-background deformation \cite{Moore:1997dj} of supersymmetric 
gauge theories provides a useful regularization procedure to study their non-perturbative 
effects \cite{Nekrasov:2002qd, Losev:2003py, Nekrasov:2003rj}. 
The $\Omega$-background is 
the curved geometry which admits the action of
the  $U(1)$ vector fields.
Since this background violates the translational symmetry,  
supersymmetry is explicitly broken.
One can however introduce the constant R-symmetry Wilson line gauge field to
retain a part of supersymmetry.

The scalar supercharge, 
which is obtained by the topological twist of supersymmetry \cite{Witten:1988ze}, 
is a particularly important ingredient in the deformed theory
because it is used to perform the path-integral exactly
via the localization formula.
For the $\Omega$-deformed ${\cal N}=2$ supersymmetric gauge theories 
\cite{Nekrasov:2002qd, Losev:2003py},
the supercharge is nilpotent up to the gauge transformation and the 
$U(1)^2$ rotation 
(see also 
\cite{Ito:2010vx} for their explicit off-shell transformations).
The 
$\Omega$-deformed $\mathcal{N} = 2$ super Yang-Mills theory 
 is also realized by the dimensional reduction of 
the ${\cal N}=1$ super Yang-Mills theory in the six-dimensional 
background \cite{Losev:2003py}. 

It is interesting to explore the ten-dimensional 
$\Omega$-background and their dimensional reduction to lower
dimensions for studying the various $\Omega$-deformed theories in a
systematic way.
In the previous paper \cite{Ito:2012hs}, we have 
studied the four-dimensional $\mathcal{N} = 4$ super Yang-Mills theory in the 
ten-dimensional $\Omega$-background 
\cite{Ito:2011cr} (see also \cite{Hellerman:2012rd, Reffert:2011dp} 
for different generalization).
Starting from the ten-dimensional $\mathcal{N} = 1$ super Yang-Mills theory in
the general curved background with torsion, we considered 
its dimensional reduction to four dimensions. 
We examined the parallel spinor conditions and the torsion 
constraints for the existence of the spinor associated with 
the scalar supercharges in the four-dimensional theory. 
The constant $SU(4)_I$ R-symmetry Wilson line gauge field, 
which is necessary to preserve supersymmetry, is identified with 
the contorsion. 
We have solved the parallel spinor conditions and the 
torsion constraints for the $\Omega$-backgrounds.
We obtained the on-shell deformed scalar supersymmetries 
associated with the three different topological twists, the half, the
Vafa-Witten and the Marcus twists \cite{Yamron, Vafa:1994tf,
Marcus:1995mq}, which were classified by Yamron \cite{Yamron}. 
These twists correspond to the theories with 
the single scalar
supercharge, the two charges with the same chirality, the two charges with
opposite chirality, respectively.

In this paper we will study further the $\Omega$-deformed ${\cal N}=4$ super
Yang-Mills theory.
We are particularly interested in the off-shell supersymmetries, which are 
deformed non-trivially in the $\Omega$-background.
We will introduce the auxiliary fields to construct 
the scalar supersymmetry acting on them. 
It will be shown that the deformed scalar supercharges form the algebra, where 
they are nilpotent and their anti-commutator vanishes up to the gauge 
transformation and the Lorentz rotation associated with the $U(1)$
vector fields.

Based on the construction of the off-shell supersymmetry algebra, we will study the 
cohomological properties of the $\Omega$-deformed action.
We will show that the deformed action
is written in the exact form with respect to the deformed scalar supercharges.
Our results show that the twisted ${\cal N}=4$ super Yang-Mills theories are also 
well-defined in the $\Omega$-backgrounds.

The organization of this paper is as follows. 
In  Section 2, we review the four-dimensional $\mathcal{N} = 4$
super Yang-Mills theory in the $\Omega$-background.
We will introduce three types of the topological twists and 
the on-shell scalar supersymmetries. 
In Section 3, we will introduce the auxiliary fields to the theory 
and study the deformed supersymmetry transformations off-shell. 
We show that the action is written in the
exact form with respect to the scalar supercharges. 
Section 4 is devoted for conclusion and discussions.
In Appendix we summarize the Dirac matrices in four and six dimensions.

\section{$\Omega$-deformed $\mathcal{N} = 4$ super Yang-Mills theory and
 on-shell supersymmetries}
In this section, we review the $\Omega$-deformation of $\mathcal{N} = 4$ super Yang-Mills
theory.
This theory is obtained by the dimensional reduction of the 
$\mathcal{N} = 1$ super Yang-Mills theory in the ten-dimensional 
spacetime with the metric \cite{Ito:2011cr}
\begin{align}
& 
ds^2_{10} = (d x^{a+4})^2 + (d x^{\mu} + \Omega_a^{\mu} d x^{a+4})^2.
\label{eq:10d_Omega}
\end{align}
Here $x^{\mu}, x^{a+4}$ ($\mu = 1, \cdots, 4, \ a = 1, \cdots, 6$) 
are the spacetime coordinates.  
$\Omega^{\mu} {}_a = \Omega^{\mu \nu} {}_a x_{\nu}$ 
are the $U(1)^6$ vector fields acting on $x^\mu$. 
The constant matrices $\Omega_{\mu \nu a}$ 
are anti-symmetric and satisfy the commutation relations:
\begin{align}
\Omega_{\mu} {}^{\rho} {}_a \Omega_{\rho \nu b} 
- \Omega_{\mu} {}^{\rho} {}_{b} \Omega_{\rho \nu a} = 0.
\end{align}
In the following we use the representation of
the matrices $\Omega_{\mu \nu a}$, which are parametrized as 
\begin{align}
\Omega_{\mu \nu a} = 
\left(
\begin{array}{cccc}
0 & \epsilon^1_{a} & 0 & 0 \\
- \epsilon^1_{a} & 0 & 0 & 0 \\
0 & 0 & 0 & - \epsilon^2_{a} \\
0 & 0 & \epsilon^2_{a} & 0
\end{array}
\right),
\label{eq:omega_matrices}
\end{align}
where $\epsilon^1_a$, $\epsilon^2_a$ are real constants.

We now compactify the $x^5, \cdots, x^{10}$ directions on the six-torus $\mathbf{T}^6$ 
and perform the dimensional reduction to four dimensions. 
The ten-dimensional Lorentz group $SO(10)$ becomes 
$SO(4) \times SO(6)_I$ where $SO(4)$ is the four-dimensional Lorentz
group and $SO(6)_I \simeq SU(4)_I$ is the R-symmetry group.
We further introduce the constant $SU(4)_I$ R-symmetry Wilson line gauge
field $(\mathcal{A}_a)^A {}_B$.
Here the index $a$ labels the vector representation of $SO(6)_I$ while 
$A$ labels the (anti)fundamental representation of $SU(4)_I$.
Then the dimensionally reduced action is given by \cite{Ito:2011cr}
\begin{align}
 S = \frac{1}{\kappa g^2} 
\int \! d^4 x \ 
\mathrm{Tr} 
\Big[ 
 &\, \frac{1}{4} F^{\mu \nu} F_{\mu \nu} 
+ \Lambda^A \sigma^{\mu} D_{\mu} \bar{\Lambda}_A + 
\frac{1}{2} \big( D_{\mu} \varphi_a -  F_{\mu \nu} \Omega^{\nu}_a \big)^2 \notag \\
 &\, - \frac{1}{2} (\Sigma_a )^{AB} \bar{\Lambda}_A [ \varphi_a ,
 \bar{\Lambda}_B ] 
- \frac{1}{2} (\bar{\Sigma}_a )_{AB} \Lambda^A [ \varphi_a , \Lambda^B ] \notag \\
 &\, - \frac{1}{4} \Big( 
[ \varphi_a , \varphi_b ] 
+ i \Omega^{\mu}_a D_{\mu} \varphi_b - i \Omega^{\mu}_b D_{\mu}
 \varphi_a 
- i F_{\mu \nu} \Omega^{\mu}_a \Omega^{\nu}_b \notag \\
 &\, \qquad \qquad - \frac{1}{2} \big( (\Sigma_b \bar{\Sigma}_c
 )^{A}{}_{B} \varphi_{c} (\mathcal{A}_a )^{B}{}_{A} - (\Sigma_a
 \bar{\Sigma}_c )^{A}{}_{B} \varphi_{c} (\mathcal{A}_b )^{B}{}_{A} \big)
 \Big)^2 \notag \\
 &\, - \frac{i}{2} \Omega^{\mu}_a \big( ( \Sigma_a )^{AB}
 \bar{\Lambda}_A 
D_{\mu} \bar{\Lambda}_B + (\bar{\Sigma}_a )_{AB} \Lambda^A D_{\mu} \Lambda^B \big) \notag \\
 &\, + \frac{i}{4} \Omega_{\mu \nu a} \big( (\Sigma_a )^{AB}
 \bar{\Lambda}_A \bar{\sigma}^{\mu \nu} \bar{\Lambda}_B 
+ (\bar{\Sigma}_a )_{AB} \Lambda^A \sigma^{\mu \nu} \Lambda^B \big) \notag \\
 &\, + \frac{1}{2} (\Sigma_a )^{AB} \bar{\Lambda}_A \bar{\Lambda}_{D}
 (\mathcal{A}_a )^{D}{}_{B} - \frac{1}{2} (\bar{\Sigma}_a )_{AB}
 \Lambda^A (\mathcal{A}_a )^{B}{}_{D} \Lambda^{D} \Big].
\label{eq:on-shell_action}
\end{align}
Here $A_{\mu} \ (\mu =1,2,3,4)$ is a gauge field, 
$\Lambda^A_{\alpha}, \ \bar{\Lambda}_{\dot{\alpha} A} \ (\alpha,
\dot{\alpha} = 1,2, A = 1,2,3,4)$ are Weyl fermions
and $\varphi_a \ (a = 1, \cdots, 6)$ are real scalar fields.
The fields are in the adjoint representation of a gauge group.
The constant $\kappa$ denotes the normalization of the Lie algebra
of the gauge group and $g$ is the gauge coupling constant.
The indices $\alpha, \ \dot{\alpha}$ represent left and right spinors of
the Lorentz group $SO(4) \simeq SU(2)_L \times SU(2)_R$. 
These indices are raised and lowered by the anti-symmetric symbols
$\epsilon_{\alpha \beta}$, $\epsilon_{\dot{\alpha} \dot{\beta}}$ with
$\epsilon^{12} = - \epsilon_{12} = 1$.
The conventions of four- and six-dimensional Dirac matrices 
$\sigma^{\mu}$, $\bar{\sigma}^{\mu}$, $(\Sigma_a)^{AB}$,
$(\bar{\Sigma}_a)_{AB}$ are given in Appendix.
The gauge covariant derivative is defined by $D_{\mu} * = \partial_{\mu} * + i
[A_{\mu}, *]$. 
The field strength of the gauge field is $F_{\mu \nu} = \partial_{\mu}
A_{\nu} - \partial_{\nu} A_{\mu} + i [A_{\mu}, A_{\nu}]$.

The action \eqref{eq:on-shell_action} is also obtained 
by replacing the following terms in the undeformed action as 
\begin{align}
& [\varphi_a, \varphi_b] \longrightarrow 
[\varphi_a , \varphi_b] 
+ i \Omega^{\mu}_a D_{\mu} \varphi_b - i \Omega^{\mu}_b D_{\mu}
 \varphi_a 
- i F_{\mu \nu} \Omega^{\mu}_a \Omega^{\nu}_b \notag \\
&\, \qquad \qquad - \frac{1}{2} \big( (\Sigma_b \bar{\Sigma}_c
 )^{A}{}_{B} \varphi_{c} (\mathcal{A}_a )^{B}{}_{A} - (\Sigma_a
 \bar{\Sigma}_c )^{A}{}_{B} \varphi_{c} (\mathcal{A}_b )^{B}{}_{A}
 \big), 
\notag \\
& D_{\mu} \varphi_a \longrightarrow D_{\mu} \varphi_a - F_{\mu \nu}
 \Omega^{\nu}_a,
\notag \\
& [\varphi_a, \Lambda^A] \longrightarrow [\varphi_a, \Lambda^A] 
{}
+ i \Omega_{a}^{\mu} D_{\mu} \Lambda^{A}
- \frac{i}{2} \Omega_{\mu \nu a} \sigma^{\mu \nu} \Lambda^A
+ (\mathcal{A}_a)^A {}_B \Lambda^B, 
\notag \\
& [\varphi_a, \bar{\Lambda}_A] \longrightarrow 
[\varphi_a, \bar{\Lambda}_A] 
{}
+ i \Omega_{a}^{\mu} D_{\mu} \bar{\Lambda}_{A}
- \frac{i}{2} \Omega_{\mu \nu a} \bar{\sigma}^{\mu \nu} \bar{\Lambda}_A
- \bar{\Lambda}_B (\mathcal{A}_a)^B {}_A.
\label{eq:shift}
\end{align}

For generic $\Omega_{\mu \nu a}$ and $(\mathcal{A}_a)^A {}_B$, 
supersymmetry of the theory is broken explicitly. 
However, a part of supersymmetry is recovered when $\Omega_{\mu \nu a}$
and $(\mathcal{A}_a)^A {}_B$ satisfy certain conditions.
When the matrices $\Omega_{\mu \nu a}$ are (anti)self-dual and
$(\mathcal{A}_a)^A {}_B = 0$, the theory has 
anti-chiral (chiral) half of the $\mathcal{N} = 4$ supersymmetry
\cite{Ito:2011cr}. 
When $\Omega_{\mu \nu a}$ are neither self-dual nor anti-self-dual, 
the supersymmetry condition can be studied by the parallel spinor
conditions in ten-dimensional $\Omega$-background with the torsion, where the
torsion is identified with the R-symmetry Wilson line gauge field.
We can solve the parallel spinor conditions and the constraints for 
the torsion in the $\Omega$-background, 
which preserve gauge symmetry in four dimensions 
and supersymmetry associated with the topological twist.
These conditions are satisfied for special $\Omega_{\mu \nu a}$ and
$(\mathcal{A}_a)^A {}_B$.
Then we obtain the topologically twisted supersymmetries 
deformed by $\Omega_{\mu \nu a}$ and $(\mathcal{A}_a)^A {}_B$ \cite{Ito:2012hs}. 

The topological twist is to 
pick an embedding of the Lorentz group $SO(4) \simeq SU(2)_L \times
SU(2)_R$ into the R-symmetry group $SU(4)_I$ 
and to define a new Lorentz group.
Let us take the $SU(2)_{L'} \times SU(2)_{R'}$ subgroup of $SU(4)_I$ 
such that the index $A$ is decomposed into $A' = 1,2$ and $\hat{A} =
3,4$. Here $A'$ and $\hat{A}$ are indices for the two-dimensional representations of 
$SU(2)_{R'}$ and $SU(2)_{L'}$, respectively. 
The vector index $a$ is also decomposed into $a'=1,2$ which 
corresponds to the two $SU(2)_{R'}\times SU(2)_{L'}$ singlets, 
and $\hat{a}=3, \cdots ,6$ which labels the $SU(2)_{R'}\times SU(2)_{L'}$
bifundamental representation.

Topological twists in $\mathcal{N} = 4$ super Yang-Mills theory are
classified into three types \cite{Yamron}.
They are called the half twist, the Vafa-Witten twist \cite{Vafa:1994tf}
and the Marcus twist \cite{Marcus:1995mq} (or the GL twist in
\cite{Kapustin:2006pk}). In the following subsections, we 
summarize the three types of topological twists and the deformed
supersymmetries.

\subsection{The half twist}
In the case of the half twist, 
we replace $SU(2)_R$ by the diagonal subgroup of $SU(2)_{R'} \times SU(2)_R$. 
The new Lorentz group becomes $SU(2)_L \times [SU(2)_{R'} \times
SU(2)_{R}]_{\mathrm{diag}}$. Here the subscript ``diag'' stands for the
diagonal subgroup which means that the spinor index $\dot{\alpha}$ 
and the R-symmetry index $A'$ are identified.
The Weyl fermions  
$\Lambda_{\alpha}^{A'}$ and $\bar{\Lambda}_{\dot{\alpha}}^{A'}$ 
are rewritten as follows,
\begin{align}
\bar{\Lambda}^{A'}_{\dot{\alpha}} =  \frac{1}{2}
 \delta^{A'}_{\dot{\alpha}} \bar{\Lambda} + \frac{1}{2}
 (\bar{\sigma}^{\mu \nu})^{A'} {}_{\dot{\alpha}} \bar{\Lambda}_{\mu
 \nu}, 
\qquad 
\Lambda^{A'}_{\alpha} =  
\frac{1}{2} \epsilon^{A' B'} (\sigma^{\mu})_{\alpha B'} \Lambda_{\mu}.
\label{eq:half_fermion_decomp}
\end{align}
We redefine the scalar field as 
\begin{align}
\varphi^{AB} = \frac{i}{\sqrt{2}} (\Sigma_a)^{AB} \varphi_a, \quad 
\bar{\varphi}_{AB} = - \frac{i}{\sqrt{2}} (\bar{\Sigma}_{a})_{AB} \varphi_a.
\end{align}
Then using the matrix representations of $(\Sigma_a)^{AB}$ and $(\bar{\Sigma}_a)_{AB}$, 
the scalar fields are decomposed as 
\begin{align}
\varphi^{AB} = 
\left(
\begin{array}{cc}
\epsilon^{A' B'} \varphi & \varphi^{A' \hat{B}} \\
\varphi^{\hat{A} B'} & - \epsilon^{\hat{A} \hat{B}} \bar{\varphi}
\end{array}
\right),
\qquad 
\bar{\varphi}_{AB} = 
\left(
\begin{array}{cc}
\epsilon_{A'B'} \bar{\varphi} & \bar{\varphi}_{A' \hat{B}} \\
\bar{\varphi}_{\hat{A} B'} & - \epsilon_{\hat{A} \hat{B}} \varphi
\end{array}
\right),
\label{eq:scalarht1}
\end{align}
where $\varphi$, $\bar{\varphi}$ are 
the $SU(2)_{R'} \times SU(2)_{L'}$ singlets and $\varphi^{A' \hat{B}}$ 
belongs to the $SU(2)_{R'} \times SU(2)_{L'}$ bifundamental representation.

The supercharges $\bar{Q}^{A'}_{\dot{\alpha}}$ and
$Q_{\alpha}^{A'}$ are decomposed into the scalar $\bar{Q}$, 
the tensor $\bar{Q}_{\mu \nu}$ and the vector $Q_\mu$.
In \cite{Ito:2012hs} we have examined the torsion 
and the parallel spinor conditions for preserving $\bar{Q}$.
The solution to the conditions is given by
\begin{align}
& \Omega_{\mu \nu a'} 
= 
\left(
\begin{array}{cccc}
0 & \epsilon^1_{a'} & 0 & 0 \\
- \epsilon^1_{a'} & 0 & 0 & 0 \\
0 & 0 & 0 & - \epsilon^2_{a'} \\
0 & 0 & \epsilon^2_{a'} & 0
\end{array}
\right), \qquad 
(\mathcal{A}_{a'})^A {}_B = 
\left(
\begin{array}{cc}
\frac{1}{4} (\epsilon_{a'}^1 + \epsilon_{a'}^2) \tau^3 & 0 \\
0 & m_{a'} \tau^3
\end{array}
\right),
 \notag \\
& 
\Omega_{\mu \nu \hat{a}} = (\mathcal{A}_{\hat{a}})^A {}_B = 0, 
\quad (a' = 1,2, \ \hat{a} = 3,4,5,6),
\label{eq:half_SUSY_condition}
\end{align}
where $m_{a'}$ are real parameters.
The Wilson line gauge fields  
$(\mathcal{A})^{\hat{A}} {}_{\hat{B}}, (\bar{\mathcal{A}})^{\hat{A}} {}_{\hat{B}}$ 
are identified with the mass matrices $M^{\hat{A}} {}_{\hat{B}} =m\tau^3$,
$\bar{M}^{\hat{A}} {}_{\hat{B}} = \bar{m}\tau^3$ 
of the adjoint hypermultiplet of the $\mathcal{N} = 2^{*}$
theory \cite{Nekrasov:2003rj, Ito:2011cr}.
Here we defined 
\begin{align}
& \mathcal{A} = \frac{1}{\sqrt{2}} (\mathcal{A}_1 - i \mathcal{A}_2), 
\qquad 
\bar{\mathcal{A}} = \frac{1}{\sqrt{2}} (\mathcal{A}_1 + i
\mathcal{A}_2),
\notag \\
& 
m = \frac{1}{\sqrt{2}} (m_1 - i m_2), 
\qquad 
\bar{m} = \frac{1}{\sqrt{2}} (m_1 + i m_2).
\end{align}
We note that the matrices
\begin{align}
\Omega_{\mu \nu} =& 
 \left(
 \begin{array}{cccc}
 0 & \epsilon^1 & 0 & 0 \\
 - \epsilon^1 & 0 & 0 & 0 \\
 0 & 0 & 0 & - \epsilon^2 \\
 0 & 0 & \epsilon^2 & 0
 \end{array}
 \right)
, \quad 
 \bar{\Omega}_{\mu \nu} = 
 \left(
 \begin{array}{cccc}
 0 & \bar{\epsilon}^1 & 0 & 0 \\
 - \bar{\epsilon}^1 & 0 & 0 & 0 \\
 0 & 0 & 0 & - \bar{\epsilon}^2 \\
 0 & 0 & \bar{\epsilon}^2 & 0
 \end{array}
 \right), 
 \notag \\
\epsilon^i =& \frac{1}{\sqrt{2}} (\epsilon^i_1 - i \epsilon^i_2), 
 \quad 
 \bar{\epsilon}^i = 
 \frac{1}{\sqrt{2}} (\epsilon^i_1 + i \epsilon^i_2),
 \quad (i=1,2)
\label{eq:N2omega}
\end{align}
characterize the $\Omega$-background defined in six dimensions
\cite{Moore:1997dj, Ito:2010vx}.
The $\bar{Q}$-transformations are given by 
\begin{align}
& \bar{Q} A_{\mu} = \Lambda_{\mu}, 
\notag 
\\
& \bar{Q} \Lambda_{\mu} = - 2 \sqrt{2} (D_{\mu} \varphi - F_{\mu \nu}
 \Omega^{\nu}), 
\notag 
\\
& \bar{Q} \varphi = \Omega^{\mu} \Lambda_{\mu}, 
\notag 
\\
& \bar{Q} \bar{\varphi} = - \sqrt{2} \bar{\Lambda} + \bar{\Omega}^{\mu}
 \Lambda_{\mu}, 
\notag 
\\
& \bar{Q} \bar{\Lambda} = - 2 i 
\left(
[\varphi, \bar{\varphi}] + i \Omega^{\mu} D_{\mu} \bar{\varphi} 
- i \bar{\Omega}^{\mu} D_{\mu} \varphi + i \bar{\Omega}^{\mu}
 \Omega^{\nu} F_{\mu \nu}
\right), 
\notag 
\\
& \bar{Q} \bar{\Lambda}_{\mu \nu} = - 2 F_{\mu \nu}^{-} 
- i (\bar{\sigma}_{\mu \nu})^{\dot{\beta}} {}_{\dot{\alpha}}
[\varphi^{\dot{\alpha}\hat{A}},\bar{\varphi}_{\hat{A}\dot{\beta}}]
, 
\notag 
\\
& \bar{Q} \varphi^{\dot{\alpha} \hat{A}} 
= - \sqrt{2} \bar{\Lambda}^{\dot{\alpha} \hat{A}}
, 
\notag 
\\
& \bar{Q} \bar{\Lambda}^{\dot{\alpha} \hat{A}} 
= - 2 i 
\left(
[\varphi, \varphi^{\dot{\alpha} \hat{A}}] 
+ i \Omega^{\mu} D_{\mu} \varphi^{\dot{\alpha} \hat{A}} 
+ M^{\hat{A}} {}_{\hat{B}} \varphi^{\dot{\alpha} \hat{B}} 
\right) 
- \Omega^{\mu \nu} (\bar{\sigma}_{\mu \nu})^{\dot{\alpha}}
 {}_{\dot{\beta}} \varphi^{\dot{\beta} \hat{A}}, 
\notag 
\\
& \bar{Q} \Lambda^{\hat{A}}_{\alpha} = 
\sqrt{2} (\sigma^{\mu})_{\alpha \dot{\alpha}} 
D_{\mu} \varphi^{\dot{\alpha} \hat{A}}.
\label{eq:half_OS}
\end{align}
Here $\Omega^{\mu}$ and $\bar{\Omega}^{\mu}$ are defined by
\begin{gather}
\Omega^{\mu} = \Omega^{\mu\nu}x_{\nu},\quad
\bar{\Omega}^{\mu} = \bar{\Omega}^{\mu\nu}x_{\nu}.
\end{gather}
The superscript $\pm$ stands for the (anti)self-dual part of a tensor 
$X_{\mu \nu}$:
\begin{align}
X^{\pm}_{\mu \nu} = \frac{1}{2} (X_{\mu \nu} \pm \tilde{X}_{\mu \nu}),
\end{align}
where $\tilde{X}_{\mu \nu} = \frac{1}{2} \epsilon_{\mu \nu \rho \sigma}
X^{\rho \sigma}$ is the dual of $X_{\mu \nu}$ and 
$\epsilon_{\mu\nu\rho\sigma}$ is the totally
antisymmetric tensor with $\epsilon_{1234}=1$.
The supercharge $\bar{Q}$ is nilpotent up to the gauge transformation, 
the Lorentz rotation associated with the $U(1) \times U(1)$ rotation
generated by $\Omega_\mu$ and the $SU(2)_{L'}$ rotation.
In fact, the nilpotency of $\bar{Q}$ on the 
fermions $\bar{\Lambda}_{\mu\nu}$ and $\Lambda_{\alpha}^{\hat{A}}$
holds by imposing the equations of motion.

\subsection{The Vafa-Witten twist}
In the case of the Vafa-Witten twist, 
we replace $SU(2)_R$ by the diagonal subgroup of $SU(2)_{L'} \times
SU(2)_{R'} \times SU(2)_R$. 
The new Lorentz group becomes $SU(2)_L \times [SU(2)_{L'} \times
SU(2)_{R'} \times SU(2)_R]_{\mathrm{diag}}$. 
The spinor index $\dot{\alpha}$ is identified with 
the R-symmetry index $A'$ and $\dot{\alpha}$ is also
identified with $\hat{A}$.
We decompose the fermion fields as
\begin{align}
& \Lambda_{\alpha A'} =  \frac{1}{2} (\sigma^{\mu})_{\alpha A'}
 \Lambda_{\mu}, \quad 
\Lambda_{\alpha \hat{A}} =  \frac{1}{2} (\sigma^{\mu})_{\alpha \hat{A}}
 \hat{\Lambda}_{\mu}, 
\notag 
\\
& \bar{\Lambda}^{A'}_{\dot{\alpha}} =  \frac{1}{2}
 \delta^{A'}_{\dot{\alpha}} \bar{\Lambda} + \frac{1}{2}
 (\bar{\sigma}^{\mu \nu})^{A'} {}_{\dot{\alpha}} \bar{\Lambda}_{\mu
 \nu}, \quad 
 \bar{\Lambda}^{\hat{A}}_{\dot{\alpha}} = \frac{1}{2}
 \delta^{\hat{A}}_{\dot{\alpha}} \hat{\bar{\Lambda}} 
+ \frac{1}{2} (\bar{\sigma}^{\mu \nu})^{\hat{A}} {}_{\dot{\alpha}}
 \hat{\bar{\Lambda}}_{\mu \nu}.
\end{align}
The scalar fields in (\ref{eq:scalarht1}) are further decomposed as
\begin{align}
&
\varphi^{A'B'} = \epsilon^{A'B'}\varphi,\quad 
\varphi^{\hat{A}\hat{B}} = -\epsilon^{\hat{A}\hat{B}}\bar{\varphi},\quad
\varphi^{A'\hat{A}} =  
\frac{1}{2} \epsilon^{A'\hat{A}} \hat{\varphi} + \frac{1}{2}
 \epsilon^{\hat{A} \hat{B}} (\bar{\sigma}^{\mu \nu})^{A'}
 {}_{\hat{B}} \hat{\varphi}_{\mu \nu}.
\label{eq:VW_fermion_decomp}
\end{align}
The supercharges $Q^A_{\alpha}$ and $\bar{Q}^{\dot{\alpha}}_A$ are
decomposed into the two scalars $\bar{Q}$, $\hat{\bar{Q}}$, the two
vectors $Q_{\mu}$, $\hat{Q}_{\mu}$ and the two tensor supercharges
$\bar{Q}_{\mu \nu}$, $\hat{\bar{Q}}_{\mu \nu}$.

The solution to the parallel spinor and the torsion conditions 
for preserving $\bar{Q}$ and $\hat{\bar{Q}}$ is  
\begin{align}
& \Omega_{\mu \nu a'} 
= 
\left(
\begin{array}{cccc}
0 & \epsilon_{a'}^1 & 0 & 0 \\
- \epsilon_{a'}^1 & 0 & 0 & 0 \\
0 & 0 & 0 & - \epsilon_{a'}^2 \\
0 & 0 & \epsilon_{a'}^2 & 0
\end{array}
\right), \quad 
\Omega_{\mu \nu \hat{a}}
= 
\left(
\begin{array}{cccc}
0 & \epsilon_{\hat{a}}^1 & 0 & 0 \\
- \epsilon_{\hat{a}}^1 & 0 & 0 & 0 \\
0 & 0 & 0 & - \epsilon_{\hat{a}}^2 \\
0 & 0 & \epsilon_{\hat{a}}^2 & 0
\end{array}
\right), 
\notag \\
& 
(\mathcal{A}_{a'})^A {}_B 
= 
\left(
\begin{array}{cc}
\frac{1}{4} (\epsilon^1_{a'} + \epsilon^2_{a'}) \tau^3
 & 0 \\
0 & 
\frac{1}{4} (\epsilon^1_{a'} + \epsilon^2_{a'}) \tau^3
\end{array}
\right), 
\quad 
(\mathcal{A}_{\hat{a}})^A {}_B 
= 
\left(
\begin{array}{cc}
\frac{1}{4} (\epsilon^1_{\hat{a}} + \epsilon^2_{\hat{a}}) \tau^3
 & 0 \\
0 & 
\frac{1}{4} (\epsilon^1_{\hat{a}} + \epsilon^2_{\hat{a}}) \tau^3
\end{array}
\right), 
\notag \\
& (a' = 1,2, \ \hat{a} = 5,6), 
\notag \\
& \Omega_{\mu \nu 3} = \Omega_{\mu \nu 4} = 
(\mathcal{A}_3)^A {}_B = (\mathcal{A}_4)^A {}_B 
= 0.
\label{eq:VW_SUSY_condition}
\end{align}
It is convenient to rewrite the $\Omega$-background matrices $\Omega_{\mu \nu a}$ as
\begin{align}
& \Omega^{AB}_{\mu \nu} = \frac{i}{\sqrt{2}} (\Sigma_a)^{AB} \Omega^a_{\mu \nu},
 \qquad 
\bar{\Omega}^{\mu \nu}_{AB} 
= - \frac{i}{\sqrt{2}} (\bar{\Sigma}_{a})_{AB} \Omega^{a \mu \nu}, 
\notag 
\\
& \Omega_{\mu}^{AB} = \Omega^{AB}_{\mu \nu} x^{\nu}, \quad 
\bar{\Omega}_{AB}^{\mu} = \bar{\Omega}_{AB}^{\mu \nu} x_{\nu}.
\end{align}
We further decompose these matrices as 
\begin{align}
& \Omega^{A'B'}_{\mu \nu} = \epsilon^{A'B'}\Omega_{\mu \nu},\quad 
\Omega^{\hat{A}\hat{B}}_{\mu \nu} = 
-\epsilon^{\hat{A}\hat{B}}\bar{\Omega}_{\mu \nu},\quad
\Omega^{A'\hat{A}}_{\mu \nu} = \frac{1}{2} \epsilon^{A'
 \hat{A}} \hat{\Omega}_{\mu \nu} + \frac{1}{2} \epsilon^{\hat{A}
 \hat{B}} (\bar{\sigma}^{\rho \sigma})^{A'} {}_{\hat{B}}
 \hat{\Omega}_{\mu \nu, \rho \sigma},
\label{eq:VW_omega_decomp}
\end{align}
where $\hat{\Omega}_{\mu \nu, \rho \sigma}$ satisfies the
anti-self-dual condition with respect to the last two indices.
From the decomposition \eqref{eq:VW_omega_decomp}, we obtain
\begin{align}
& \hat{\Omega}_{\mu \nu} = 
\sqrt{2}
\left(
\begin{array}{cccc}
0 & \epsilon^1_{6} & 0 & 0 \\
- \epsilon^1_{6} & 0 & 0 & 0 \\
0 & 0 & 0 & - \epsilon^2_{6} \\
0 & 0 & \epsilon^2_{6} & 0
\end{array}
\right)
, \quad 
\hat{\Omega}_{\mu \nu, 12} = - 
\hat{\Omega}_{\mu \nu, 34} = 
- \frac{1}{\sqrt{2}}
\left(
\begin{array}{cccc}
0 & \epsilon^1_{5} & 0 & 0 \\
- \epsilon^1_{5} & 0 & 0 & 0 \\
0 & 0 & 0 & - \epsilon^2_{5} \\
0 & 0 & \epsilon^2_{5} & 0
\end{array}
\right) 
\notag \\
& \hat{\Omega}_{\mu \nu, \rho \sigma} = 0, \quad ( (\rho, \sigma) \not= (1,2), (3,4)).
\label{eq:VW_omega_condition2}
\end{align}
The matrices $\Omega_{\mu \nu}$, $\bar{\Omega}_{\mu \nu}$ are given in \eqref{eq:N2omega}.
The $\bar{Q}$-transformations are given by 
\begin{align}
& \bar{Q} A_{\mu} = \Lambda_{\mu}, \quad
\bar{Q} \Lambda_{\mu} = - 2 \sqrt{2} 
(D_{\mu} \varphi - F_{\mu \nu} \Omega^{\nu}), 
\notag 
\\
& \bar{Q} \varphi = \Omega^{\mu} \Lambda_{\mu}, 
\notag 
\\
& \bar{Q} \bar{\varphi} =  - \sqrt{2} \bar{\Lambda} +
 \bar{\Omega}^{\mu} \Lambda_{\mu}, \quad 
\bar{Q} \bar{\Lambda} = - 2 i 
\left(
[\varphi, \bar{\varphi}] + i \Omega^{\mu} D_{\mu} \bar{\varphi} - i
 \bar{\Omega}^{\mu} D_{\mu} \varphi + i \bar{\Omega}^{\mu} \Omega^{\nu}
 F_{\mu \nu}
\right), 
\notag 
\\
& \bar{Q} \bar{\Lambda}_{\mu \nu} =  - 2 F^{-}_{\mu \nu} + i
\bigl([\hat{\varphi},\hat{\varphi}_{\mu\nu}]
+i\hat{\Omega}^{\rho}D_{\rho}\hat{\varphi}_{\mu\nu}
-i\hat{\Omega}^{\rho,}{}_{\mu\nu}D_{\rho}\hat{\varphi}
+i\hat{\Omega}^{\rho,}{}_{\mu\nu}\hat{\Omega}^{\sigma}F_{\rho\sigma}
-i\hat{\Omega}_{\mu}{}^{\rho}\hat{\varphi}_{\rho\nu}
+i\hat{\Omega}_{\nu}{}^{\rho}\hat{\varphi}_{\rho\mu}\bigr)
\notag\\
& \qquad\qquad - \frac{i}{2}
\bigl([\hat{\varphi}_{\mu\rho},\hat{\varphi}_{\nu}{}^{\rho}]
+i\hat{\Omega}^{\rho,}{}_{\mu\sigma}D_{\rho}\hat{\varphi}_{\nu}{}^{\sigma}
-i\hat{\Omega}^{\rho,}{}_{\nu\sigma}D_{\rho}\hat{\varphi}_{\mu}{}^{\sigma}
+\hat{\Omega}^{-}_{\mu\nu,\rho\sigma} \hat{\varphi}^{\rho\sigma}
-\hat{\Omega}_{\rho\sigma}{}^{\rho\sigma}\hat{\varphi}_{\mu\nu}\bigr)
,
\notag 
\\
& \bar{Q} \hat{\Lambda}_{\mu} =  - \sqrt{2} (D_{\mu} \hat{\varphi} - F_{\mu \nu} \hat{\Omega}^{\nu}) - 2 \sqrt{2} (D^{\nu} \hat{\varphi}_{\mu \nu} - F^{\nu \rho} \hat{\Omega}_{\rho, \mu \nu})
, 
\notag 
\\
& \bar{Q} \hat{\varphi} = - \sqrt{2} \hat{\bar{\Lambda}} +
 \hat{\Omega}^{\mu} \Lambda_{\mu}, \quad 
\bar{Q} \hat{\bar{\Lambda}} = - 2 i 
\left(
[\varphi, \hat{\varphi}] + i \Omega^{\mu} D_{\mu} \hat{\varphi} - i
 \hat{\Omega}^{\mu} D_{\mu} \varphi + i \hat{\Omega}^{\mu} \Omega^{\nu}
 F_{\mu \nu}
\right), 
\notag 
\\
& \bar{Q} \hat{\varphi}_{\mu \nu} = - \sqrt{2}
 \hat{\bar{\Lambda}}_{\mu \nu} + \hat{\Omega}^{\rho,} {}_{\mu \nu}
 \Lambda_{\rho}, 
\notag \\
& \bar{Q} \hat{\bar{\Lambda}}_{\mu \nu} = 
- 2 i 
\left(
[\varphi, \hat{\varphi}_{\mu \nu}] + i \Omega^{\rho} D_{\rho}
 \hat{\varphi}_{\mu \nu} - i \hat{\Omega}^{\rho,} {}_{\mu \nu} D_{\rho}
 \varphi + i \hat{\Omega}^{\rho,} {}_{\mu \nu} \Omega^{\sigma} F_{\rho \sigma}
- i \Omega_{\mu} {}^{\rho} \hat{\varphi}_{\rho \nu} + i \Omega_{\nu} {}^{\rho}
 \hat{\varphi}_{\rho \mu}
\right).
\label{eq:VW_OS1}
\end{align}
Here $\hat{\Omega}^{\mu}$ and $\hat{\Omega}^{\rho,}{}_{\mu\nu}$ 
are defined by
\begin{gather}
\hat{\Omega}^{\mu} = \hat{\Omega}^{\mu\nu}x_{\nu},\quad
\hat{\Omega}^{\rho,}{}_{\mu\nu} = 
\hat{\Omega}^{\rho\sigma,}{}_{\mu\nu}\,x_{\sigma}, 
\end{gather}
and $\hat{\Omega}^{-}_{\mu\nu,\rho\sigma}$ is the anti-self-dual part of 
$\hat{\Omega}_{\mu\nu,\rho\sigma}$ with respect to the first two indices. 
The $\hat{\bar{Q}}$-transformations are also given by  
\begin{align}
& \hat{\bar{Q}} A_{\mu}
=
\hat{\Lambda}_{\mu},
\quad 
\hat{\bar{Q}} \hat{\Lambda}_{\mu}
=
2\sqrt{2}(D_{\mu}\bar{\varphi}-F_{\mu\nu}\bar{\Omega}^{\nu}),
\notag \\
& \hat{\bar{Q}} \bar{\varphi}
=
\bar{\Omega}^{\mu}\hat{\Lambda}_{\mu},
\notag \\
& \hat{\bar{Q}} \varphi
=
\sqrt{2}\hat{\bar{\Lambda}}+\Omega^{\mu}\hat{\Lambda}_{\mu},
\quad 
\hat{\bar{Q}} \hat{\bar{\Lambda}}
=
2i([\varphi,\bar{\varphi}]+i\Omega^{\mu}D_{\mu}\bar{\varphi}
-i\bar{\Omega}^{\mu}D_{\mu}\varphi+i\bar{\Omega}^{\mu}\Omega^{\nu}F_{\mu\nu}),
\notag \\
& \hat{\bar{Q}} \hat{\bar{\Lambda}}_{\mu\nu}
= - 2 F^{-}_{\mu \nu}
-i([\hat{\varphi},\hat{\varphi}_{\mu\nu}]
+i\hat{\Omega}^{\rho}D_{\rho}\hat{\varphi}_{\mu\nu}
-i\hat{\Omega}^{\rho,}{}_{\mu\nu}D_{\rho}\hat{\varphi}
+i\hat{\Omega}^{\rho,}{}_{\mu\nu}\hat{\Omega}^{\sigma}F_{\rho\sigma}
-i\hat{\Omega}_{\mu}{}^{\rho}\hat{\varphi}_{\rho\nu}
+i\hat{\Omega}_{\nu}{}^{\rho}\hat{\varphi}_{\rho\mu}), \notag \\
& \qquad\qquad - \frac{i}{2}
\bigl([\hat{\varphi}_{\mu\rho},\hat{\varphi}_{\nu}{}^{\rho}]
+i\hat{\Omega}^{\rho,}{}_{\mu\sigma}D_{\rho}\hat{\varphi}_{\nu}{}^{\sigma}
-i\hat{\Omega}^{\rho,}{}_{\nu\sigma}D_{\rho}\hat{\varphi}_{\mu}{}^{\sigma}
+\hat{\Omega}^{-}_{\mu\nu,\rho\sigma}\hat{\varphi}^{\rho\sigma}
-\hat{\Omega}_{\rho\sigma,}{}^{\rho\sigma}\hat{\varphi}_{\mu\nu}\bigr),
 \notag \\
& \hat{\bar{Q}} \Lambda_{\mu}
= - \sqrt{2} (D_{\mu} \hat{\varphi} - F_{\mu \nu} \hat{\Omega}^{\nu})
+ 2 \sqrt{2} (D^{\nu} \hat{\varphi}_{\mu \nu} - F^{\nu \rho} \hat{\Omega}_{\rho, \mu \nu})
,
 \notag \\
& \hat{\bar{Q}} \hat{\varphi}
=
\sqrt{2}\bar{\Lambda}+\hat{\Omega}^{\mu}\hat{\Lambda}_{\mu}, 
\qquad 
\hat{\bar{Q}} \bar{\Lambda}
=
-2i([\bar{\varphi},\hat{\varphi}]
+i\bar{\Omega}^{\mu}D_{\mu}\hat{\varphi}
-i\hat{\Omega}^{\mu}D_{\mu}\bar{\varphi}
+i\hat{\Omega}^{\mu}\bar{\Omega}^{\nu}F_{\mu\nu}),
\notag \\
& \hat{\bar{Q}} \hat{\varphi}_{\mu\nu}
=
\sqrt{2}\bar{\Lambda}_{\mu\nu}
+\hat{\Omega}^{\rho,}{}_{\mu\nu}\hat{\Lambda}_{\rho}, 
\notag \\
& \hat{\bar{Q}} \hat{\bar{\Lambda}}_{\mu\nu}
=
-2i([\bar{\varphi},\hat{\varphi}_{\mu\nu}]
+i\bar{\Omega}^{\rho}D_{\rho}\hat{\varphi}_{\mu\nu}
-i\hat{\Omega}^{\rho,}{}_{\mu\nu}D_{\rho}\bar{\varphi}
+i\hat{\Omega}^{\rho,}{}_{\mu\nu}\bar{\Omega}^{\sigma}F_{\rho\sigma}
-i\bar{\Omega}_{\mu}{}^{\rho}\hat{\varphi}_{\rho\nu}
+i\bar{\Omega}_{\nu}{}^{\rho}\hat{\varphi}_{\rho\mu}).
\label{eq:VW_OS2}
\end{align}
We note that when $\epsilon^{1}_{\hat{a}}=\epsilon^2_{\hat{a}}=0$ ($\hat{a}=5,6$), the
theory reduces to the $\Omega$-deformed ${\cal N}=2^*$ theory with the
mass parameters 
$m=\frac{1}{4}(\epsilon^1+\epsilon^2)$ and
$\bar{m}=
\frac{1}{4}(\bar{\epsilon}^1+\bar{\epsilon}^2)$.
In other words, we have a supersymmetry enhancement 
in the $\Omega$-deformed ${\cal N}=2^*$ theory by choosing the mass parameters 
to the above special values. 
A similar enhancement of supersymmetry was 
also discussed in \cite{Pestun:2007rz, Okuda:2010ke}.

\subsection{The Marcus twist}
In the case of the Marcus twist, 
we replace $SU(2)_L$ by the diagonal subgroup of $SU(2)_{L'} \times
SU(2)_{L}$ and $SU(2)_R$ by the diagonal subgroup of 
$SU(2)_{R'} \times SU(2)_R$. 
The new Lorentz group becomes $[SU(2)_{L'} \times
SU(2)_{L}]_{\mathrm{diag}} \times [SU(2)_{R'} \times
SU(2)_{R}]_{\mathrm{diag}}$, where 
the indices $\alpha$ and $\hat{A}$, $\dot{\alpha}$
and $A'$ are identified respectively.
We decompose the Weyl fermions as 
\begin{align}
& \Lambda^{A'}_{\alpha}
=\frac{1}{2}\epsilon^{A'\dot{\beta}}(\sigma^{\mu})_{\alpha\dot{\beta}}
\Lambda_{\mu}, \quad 
\bar{\Lambda}_{A'}^{\dot{\alpha}}
=-\frac{1}{2}\delta_{A'}^{\dot{\alpha}}\bar{\Lambda}
+\frac{1}{2}(\bar{\sigma}^{\mu\nu})^{\dot{\alpha}}{}_{A'}
\bar{\Lambda}_{\mu\nu}, 
\notag 
\\
& \bar{\Lambda}^{\dot{\alpha} \hat{A}} = \frac{1}{2} 
(\bar{\sigma}^{\mu})^{\dot{\alpha} \hat{A}} \bar{\Lambda}_{\mu}, \quad 
\Lambda_{\alpha}^{\hat{A}} = \frac{1}{2} \delta^{\hat{A}}_{\alpha} \Lambda
+ \frac{1}{2} (\sigma^{\mu \nu})_{\alpha} {}^{\hat{A}} \Lambda_{\mu
 \nu}.
\label{eq:Marcus_fermion_decomp}
\end{align}
We define the field $\varphi_{\mu}$ as 
\begin{align}
\varphi_{\mu}= (\sigma_{\mu})_{\hat{B}A'}\varphi^{A'\hat{B}}.
\end{align}
The supercharges are decomposed into the two scalars $Q$ and $\bar{Q}$,
the two vectors $Q_{\mu}$, $\bar{Q}_{\mu}$ and the two tensor
supercharges $Q_{\mu \nu}$, $\bar{Q}_{\mu\nu}$.
The solution to the parallel spinor and the torsion conditions 
for preserving $Q$ and $\bar{Q}$ is given by 
\begin{align}
& \Omega_{\mu \nu a'} 
= 
\left(
\begin{array}{cccc}
0 & \epsilon_{a'}^1 & 0 & 0 \\
- \epsilon_{a'}^1 & 0 & 0 & 0 \\
0 & 0 & 0 & - \epsilon_{a'}^2 \\
0 & 0 & \epsilon_{a'}^2 & 0
\end{array}
\right), \quad 
(\mathcal{A}_{a'})^{A} {}_{B} 
= 
\left(
\begin{array}{cc}
\frac{1}{4} (\epsilon_{a'}^1 + \epsilon_{a'}^2) \tau^3
 & 0 \\
0 & 
\frac{1}{4} (\epsilon_{a'}^1 - \epsilon_{a'}^2) \tau^3 
\end{array}
\right),
\notag \\
& 
\Omega_{\mu \nu \hat{a}} = 
(\mathcal{A}_{\hat{a}})^{A} {}_B = 0, 
\quad 
(a'=1,2, \ \hat{a} = 3,4,5,6).
\label{eq:Marcus_SUSY_condition}
\end{align}
The supersymmetry transformations generated by $\bar{Q}$ and $Q$ are 
\begin{align}
&\bar{Q} A_{\mu}
=
\Lambda_{\mu},
\qquad 
\bar{Q}\Lambda_{\mu}
=
-2\sqrt{2}(D_{\mu}\varphi-F_{\mu\nu}\Omega^{\nu}),
\notag
\\
& \bar{Q} \varphi
=
\Omega^{\mu}\Lambda_{\mu},
\notag
\\
& \bar{Q} \bar{\varphi}
=
-\sqrt{2}\bar{\Lambda}+\bar{\Omega}^{\mu}\Lambda_{\mu},
\qquad 
\bar{Q} \bar{\Lambda}
= 
-2i([\varphi,\bar{\varphi}]+i\Omega^{\mu}D_{\mu}\bar{\varphi}
-i\bar{\Omega}^{\mu}D_{\mu}\varphi+i\bar{\Omega}^{\mu}\Omega^{\nu}F_{\mu\nu}),
\notag
\\
& \bar{Q} \varphi_{\mu}
=
-\sqrt{2}\bar{\Lambda}_{\mu}, 
\qquad 
\bar{Q} \bar{\Lambda}_{\mu}
=
-2i([\varphi,\varphi_{\mu}]
+i\Omega^{\nu}D_{\nu}\varphi_{\mu}-i\Omega_{\mu}{}^{\nu}\varphi_{\nu}),
\notag
\\
& \bar{Q} \Lambda
= \sqrt{2} D_{\mu} \varphi^{\mu},  
\qquad 
\bar{Q} \Lambda_{\mu\nu}
= \sqrt{2} (D_{\mu} \varphi_{\nu} - D_{\nu} \varphi_{\mu})^{+},
\notag
\\
& \bar{Q} \bar{\Lambda}_{\mu\nu}
= - 2 F^{-}_{\mu \nu} + i [\varphi_{\mu}, \varphi_{\nu}]^{-}, 
\label{eq:Marcus_OS1}
\end{align}
and 
\begin{align}
& Q A_{\mu}
=
\bar{\Lambda}_{\mu},
\qquad 
Q \bar{\Lambda}_{\mu}
=
-2\sqrt{2}(D_{\mu}\varphi-F_{\mu\nu}\Omega^{\nu}),
\notag
\\
& Q \varphi
=
\Omega^{\mu}\bar{\Lambda}_{\mu},
\notag
\\
& Q \bar{\varphi}
=
\sqrt{2}\Lambda+\bar{\Omega}^{\mu}\bar{\Lambda}_{\mu},
\qquad 
Q \Lambda
=
2i([\varphi,\bar{\varphi}]+i\Omega^{\mu}D_{\mu}\bar{\varphi}
-i\bar{\Omega}^{\mu}D_{\mu}\varphi+i\bar{\Omega}^{\mu}\Omega^{\nu}F_{\mu\nu}),
\notag
\\
& Q \varphi_{\mu}
=
\sqrt{2}\Lambda_{\mu}, 
\qquad 
Q \Lambda_{\mu}
=
2i([\varphi,\varphi_{\mu}]
+i\Omega^{\nu}D_{\nu}\varphi_{\mu}-i\Omega_{\mu}{}^{\nu}\varphi_{\nu}),
\notag
\\
& Q \bar{\Lambda}
= \sqrt{2} D_{\mu} \varphi^{\mu}, 
\qquad 
Q \Lambda_{\mu\nu}
= - 2 F^{+}_{\mu \nu} + i [\varphi_{\mu}, \varphi_{\nu}]^{+}, 
\notag
\\
& Q \bar{\Lambda}_{\mu\nu}
=
- \sqrt{2} (D_{\mu} \varphi_{\nu} - D_{\nu} \varphi_{\mu})^{-}.
\label{eq:Marcus_OS2}
\end{align}

The $\Omega$-deformed theory corresponding to the Marcus twist can
be obtained by choosing the mass parameters 
$m=\frac{1}{4}(\epsilon^1-\epsilon^2)$ and
$\bar{m}=\frac{1}{4}(\bar{\epsilon}^1-\bar{\epsilon}^2)$ 
in the $\Omega$-deformed ${\cal N}=2^*$ theory.
The relations among three types of the $\Omega$-deformed theories are 
summarized in Fig. 1.
The $\Omega$-deformed ${\cal N}=2^*$ theory 
with generic $\epsilon^1$ and $\epsilon^2$ has two special supersymmetry 
enhancements at the mass parameters 
$m_{a'}=\frac{1}{4}(\epsilon^1_{a'}\pm \epsilon^2_{a'})$.

\begin{figure}[hbtp]
\begin{gather*}
\xymatrix{
*++[F]\txt{the half twist} 
\ar[ddd]
_{\displaystyle{m_{a'}=\frac{1}{4}(\epsilon^{1}_{a'} - \epsilon^{2}_{a'})}}
\ar[dddrr]^{\quad 
\displaystyle{m_{a'}=\frac{1}{4}(\epsilon^{1}_{a'} + \epsilon^{2}_{a'})}} 
& & 
*++[F]\txt{the Vafa-Witten twist} 
\ar[ddd]^{\displaystyle{\epsilon^{1}_{\hat{a}}=\epsilon^{2}_{\hat{a}}=0}}
\\
 & &  \\
 & &  \\
*++[F]\txt{the Marcus twist} & & 
*++[F]\txt{the half twist with 
$m_{a'}=\frac{1}{4}(\epsilon^{1}_{a'} + \epsilon^{2}_{a'})$ \\
$=$ the Vafa-Witten twist with $\epsilon^{1}_{\hat{a}}=\epsilon^{2}_{\hat{a}}=0$}
}
\end{gather*}
\caption{The relations among the theories with the different twists.}
\end{figure}
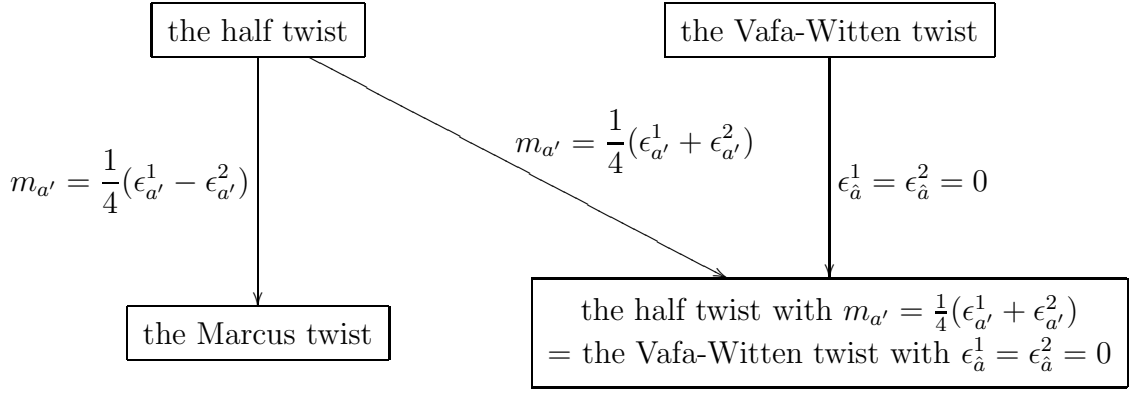

\section{Off-shell supersymmetries and exactness of action}
In order to compute the partition function and the correlation
functions of observables with the help of
localization technique, we need to investigate off-shell supersymmetry 
\cite{Pestun:2007rz}.
In this section, we study off-shell 
supersymmetry generated by the scalar supercharges in 
the $\Omega$-deformed $\mathcal{N} = 4$ super Yang-Mills theories by
introducing the auxiliary fields associated with the fermions.
For the undeformed case, the off-shell supersymmetry has been discussed
in \cite{Marcus:1995mq, Labastida:1997xk, Lozano:1999ji, Kapustin:2006pk}.

\subsection{The half twist}
We begin with the half twist case.
In the background \eqref{eq:half_SUSY_condition}, the action
\eqref{eq:on-shell_action} is invariant under the
deformed on-shell $\bar{Q}$-transformations \eqref{eq:half_OS}.
In order to get the off-shell $\bar{Q}$-supersymmetry, we introduce the auxiliary
fields $D_{\mu \nu}$ and $K^{\hat{A}}_{\alpha}$ associated with the
fermions $\bar{\Lambda}_{\mu \nu}$ and $\Lambda^{\hat{A}}_{\alpha}$.
We add the Gaussian terms of these fields to the action.
Then the action becomes
\begin{align}
S_1 = S + \frac{1}{\kappa g^2} \int \! d^4 x \
\mathrm{Tr}
\left[
- \frac{1}{2} (D_{\mu \nu})^2 
{} + \frac{1}{2} K^{\hat{A}}_{\alpha}K_{\hat{A}}^{\alpha}
\right].
\label{eq:HOS_action}
\end{align}
Now we modify the supersymmetry transformation of the fields $\bar{\Lambda}_{\mu \nu}$
and $\Lambda_{\alpha}^{\hat{A}}$ as 
\begin{align}
\bar{Q} \bar{\Lambda}_{\mu \nu} &= 2 D_{\mu \nu} - 2 F^{-}_{\mu \nu} 
- i (\bar{\sigma}_{\mu \nu})^{\dot{\beta}} {}_{\dot{\alpha}}
[\varphi^{\dot{\alpha}\hat{A}},\bar{\varphi}_{\hat{A}\dot{\beta}}], 
\notag 
\\
\bar{Q} \Lambda^{\hat{A}}_{\alpha} &= 2 K^{\hat{A}}_{\alpha} + \sqrt{2}
 (\sigma^{\mu})_{\alpha \dot{\alpha}} D_{\mu} \varphi^{\dot{\alpha} \hat{A}}.
\label{eq:half_modified_transformation}
\end{align}
Then the supersymmetry transformations of the fields $D_{\mu \nu}$ and
$K^{\hat{A}}_{\alpha}$ are determined
from the condition that the linear term in the auxiliary fields 
in the $\bar{Q}$-transformation of the Lagrangian 
vanishes. The result is
\begin{align}
\bar{Q}D_{\mu\nu}
&=
(D_{\mu}\Lambda_{\nu}-D_{\nu}\Lambda_{\mu})^{-}
-\sqrt{2}i(\bar{\sigma}_{\mu\nu})^{\dot{\beta}}{}_{\dot{\alpha}}
[\bar{\Lambda}^{\dot{\alpha}\hat{A}},\bar{\varphi}_{\hat{A}\dot{\beta}}]
\notag\\
&\quad{}
+\sqrt{2}i[\varphi,\bar{\Lambda}_{\mu\nu}]
-\sqrt{2}\Omega^{\rho}D_{\rho}\bar{\Lambda}_{\mu\nu}
+\sqrt{2}\left(
\Omega_{\mu} {}^{\rho} \bar{\Lambda}_{\rho \nu} 
- \Omega_{\nu} {}^{\rho} \bar{\Lambda}_{\rho \mu}
\right),
\notag\\[2mm]
\bar{Q}K^{\hat{A}}_{\alpha}
&=
(\sigma^{\mu})_{\alpha\dot{\alpha}}D_{\mu}\bar{\Lambda}^{\dot{\alpha}\hat{A}}
-\frac{i}{\sqrt{2}} (\sigma^{\mu})_{\alpha\dot{\alpha}}
[\Lambda_{\mu},\varphi^{\dot{\alpha}\hat{A}}]
\notag\\
&\quad{}
+\sqrt{2} i 
\left(
[\varphi, \Lambda^{\hat{A}}_{\alpha}] 
+ i \Omega^{\mu} D_{\mu} \Lambda^{\hat{A}}_{\alpha} 
+ M^{\hat{A}} {}_{\hat{B}} \Lambda^{\hat{B}} {}_{\alpha}
- \frac{i}{2} \Omega^{\mu \nu} (\sigma_{\mu \nu})_{\alpha}
 {}^{\beta} \Lambda_{\beta}^{\hat{A}}
\right).
\end{align}
Then the action \eqref{eq:HOS_action} is invariant under the $\bar{Q}$-transformation.
We redefine the auxiliary fields as 
\begin{align}
H_{\mu \nu} =& D_{\mu \nu} - F^{-}_{\mu \nu} - \frac{i}{2}
 (\bar{\sigma}_{\mu \nu})^{\dot{\beta}} {}_{\dot{\alpha}}
[\varphi^{\dot{\alpha}\hat{A}},\bar{\varphi}_{\hat{A}\dot{\beta}}],
\notag 
\\
G_{\alpha}^{\hat{A}} =& K_{\alpha}^{\hat{A}} 
+ \frac{1}{\sqrt{2}} (\sigma^{\mu})_{\alpha \dot{\alpha}} D_{\mu}
 \varphi^{\dot{\alpha} \hat{A}},
\end{align}
so that \eqref{eq:half_modified_transformation} 
takes a simple form
\begin{align}
\bar{Q} \bar{\Lambda}_{\mu \nu}&=2H_{\mu \nu}, & 
\bar{Q} \Lambda^{\hat{A}}_{\alpha}&=2G^{\hat{A}}_{\alpha}.
\label{eq:HT_aux}
\end{align}
Then the transformations for $H_{\mu \nu}$ and $G^{\hat{A}}_{\alpha}$ 
become 
\begin{align}
\bar{Q} H_{\mu \nu} &= \sqrt{2} i [\varphi, \bar{\Lambda}_{\mu \nu}] 
- \sqrt{2} \Omega^{\rho} D_{\rho} \bar{\Lambda}_{\mu \nu} 
+ \sqrt{2}
\left(
\Omega_{\mu} {}^{\rho} \bar{\Lambda}_{\rho \nu} 
- \Omega_{\nu} {}^{\rho} \bar{\Lambda}_{\rho \mu}
\right), 
\notag 
\\
\bar{Q} G^{\hat{A}}_{\alpha} &= 
\sqrt{2} i 
\left(
[\varphi, \Lambda^{\hat{A}}_{\alpha}] 
+ i \Omega^{\mu} D_{\mu} \Lambda^{\hat{A}}_{\alpha} 
+ M^{\hat{A}} {}_{\hat{B}} \Lambda^{\hat{B}} {}_{\alpha}
- \frac{i}{2} \Omega^{\mu \nu} (\sigma_{\mu \nu})_{\alpha}
 {}^{\beta} \Lambda_{\beta}^{\hat{A}}
\right).
\label{eq:half_auxiliary_transf}
\end{align}
We find that the $\bar{Q}^2$ action on a field $\Psi$
results in the form,
\begin{align}
\bar{Q}^2 \Psi
= 2 \sqrt{2} \bigl(\,\delta_{\mathrm{gauge}} (\varphi)
+ 
\delta_{\mathrm{Lorentz}} (\Omega) 
+ 
\delta_{\mathrm{flavor}} (M) \,\bigr) \Psi.
\end{align}
Here the symbol $\delta_{\mathrm{gauge}}(\varphi)$
stands for the gauge transformation by the parameter $\varphi$,
$\delta_{\mathrm{Lorentz}} (\Omega)$ 
for the Lorentz transformation by the parameter $\Omega_{\mu \nu}$ 
and $\delta_{\mathrm{flavor}} (M)$ 
for the $SU(2)_{L'}$ rotation  
by the parameter $ M^{\hat{A}} {}_{\hat{B}}$ 
defined in the Subsection 2.1.
We note that due to the conditions \eqref{eq:half_SUSY_condition} 
for the parameters, the rotations $\delta_{\mathrm{Lorentz}} (\Omega)$ and 
$\delta_{\mathrm{flavor}} (M)$ are reduced to the ones generated by their 
Cartan subgroups.
The algebra of the symmetry generated by $\bar{Q}$ closes off-shell.

We next examine the $\bar{Q}$-exactness of the action
\eqref{eq:HOS_action}.
This property is important to study the cohomological structure of the theory.
Using the transformations  
\eqref{eq:HT_aux}, 
\eqref{eq:half_auxiliary_transf}
and \eqref{eq:half_OS} for the other fields, we find that 
the action \eqref{eq:HOS_action} is written in the $\bar{Q}$-exact form:
\begin{align}
S_1 = \bar{Q} \Xi_1 + \int\! d^4 x \frac{1}{\kappa g^2} \mathrm{Tr}
\left[
\frac{1}{4} F_{\mu \nu} \tilde{F}^{\mu \nu}
\right],
\label{eq:half_exact_action}
\end{align}
where $\Xi_1$, which is called as the gauge fermion, is defined by 
\begin{align}
\Xi_1 & =  \int\! d^4 x 
\frac{1}{\kappa g^2} 
\mathrm{Tr}
\left[
- \frac{1}{2} F^{-}_{\mu \nu} \bar{\Lambda}^{\mu \nu} 
- \frac{1}{4} H_{\mu \nu} \bar{\Lambda}^{\mu \nu} 
- \frac{1}{2 \sqrt{2}} \Lambda^{\mu}
\left(
D_{\mu} \bar{\varphi} - F_{\mu \nu} \bar{\Omega}^{\nu}
\right)
\right. 
\notag \\
& \quad
+ \frac{i}{4} \bar{\Lambda}
\left(
[\varphi, \bar{\varphi}] 
+ i \Omega^{\mu} D_{\mu} \bar{\varphi} 
- i \bar{\Omega}^{\mu} D_{\mu} \varphi 
+ i \bar{\Omega}^{\mu} \Omega^{\nu} F_{\mu \nu}
\right)
\notag \\
& \quad
+ \frac{1}{2} \Lambda^{\alpha}_{\hat{A}} G^{\hat{A}}_{\alpha} 
- \frac{1}{\sqrt{2}} \Lambda^{\alpha}_{\hat{A}} (\sigma^{\mu})_{\alpha
 \dot{\alpha}} 
D_{\mu} \varphi^{\dot{\alpha} \hat{A}} 
\notag \\
& \quad
- \frac{i}{2} \bar{\Lambda}^{\dot{\alpha} \hat{A}}
\left(
[\bar{\varphi}, \bar{\varphi}_{\hat{A} \dot{\alpha}}] 
+ i \bar{\Omega}^{\mu} D_{\mu} \bar{\varphi}_{\hat{A} \dot{\alpha}} 
- \frac{i}{2} \bar{\Omega}_{\mu \nu} (\bar{\sigma}^{\mu
 \nu})^{\dot{\beta}} {}_{\dot{\alpha}} \bar{\varphi}_{\hat{A} \dot{\beta}} 
+ 
\bar{M}^{\hat{B}} {}_{\hat{A}} \bar{\varphi}_{\hat{B} \dot{\alpha}}
\right)
\notag \\
& \quad \left.
- \frac{i}{4} \bar{\Lambda}^{\mu \nu} (\bar{\sigma}_{\mu
 \nu})^{\dot{\beta}} {}_{\dot{\alpha}} 
[\varphi^{\dot{\alpha} \hat{A}}, \bar{\varphi}_{\hat{A} \dot{\beta}}] 
\right].
\label{eq:ht_fermion}
\end{align}
In \cite{Ito:2010vx}, we have shown that the $\mathcal{N} = 2$ super
Yang-Mills theory in the $\Omega$-background defined in six dimensions 
is written in the $\bar{Q}$-exact form.
When the $\mathcal{N} =2$ hypermultiplet 
($\varphi^{\dot{\alpha} \hat{A}}, \Lambda^{\alpha} {}_{\hat{A}},
\bar{\Lambda}^{\dot{\alpha} \hat{A}}, K^{\hat{A}}_{\alpha}$) is removed, 
the action \eqref{eq:half_exact_action} becomes that of the 
$\Omega$-deformed $\mathcal{N} = 2$ super Yang-Mills theory.
Then the gauge fermion \eqref{eq:ht_fermion} indeed becomes the one which was
found in the $\Omega$-deformed $\mathcal{N} = 2$ super Yang-Mills theory.

\subsection{The Vafa-Witten twist}
In the case of the Vafa-Witten twist, 
we have obtained the two on-shell scalar supercharges $\bar{Q}$ and
$\hat{\bar{Q}}$ with the same chirality in Subsection 2.2. 
The on-shell transformations by $\bar{Q}$, $\hat{\bar{Q}}$ 
are \eqref{eq:VW_OS1} and \eqref{eq:VW_OS2}.
For the undeformed case, the off-shell supersymmetry transformations 
were constructed in \cite{Yamron,Labastida:1997xk}.

We first consider the deformed off-shell supersymmetry generated by $\bar{Q}$.
Following \cite{Labastida:1997xk}, we introduce the auxiliary fields 
$D_{\mu \nu}$, $K_{\mu}$.
Then we add the quadratic terms of the auxiliary fields to 
the action \eqref{eq:on-shell_action} as 
\begin{align}
S_2 = S + \int d^{4}x 
\frac{1}{\kappa g^2} \mathrm{Tr}
\left[
- \frac{1}{2} (D_{\mu \nu})^2 - \frac{1}{2} (K_{\mu})^2
\right].
\label{eq:VW_action}
\end{align}
We modify the $\bar{Q}$-transformations of the fields 
$\bar{\Lambda}_{\mu \nu}$, $\hat{\Lambda}_{\mu}$ as 
\begin{align}
\bar{Q} \bar{\Lambda}_{\mu \nu} & =  2 D_{\mu \nu} - 2 F^{-}_{\mu \nu} 
\notag\\
& \quad
+ i \bigl([\hat{\varphi},\hat{\varphi}_{\mu\nu}]
+i\hat{\Omega}^{\rho}D_{\rho}\hat{\varphi}_{\mu\nu}
-i\hat{\Omega}^{\rho,}{}_{\mu\nu}D_{\rho}\hat{\varphi}
+i\hat{\Omega}^{\rho,}{}_{\mu\nu}\hat{\Omega}^{\sigma}F_{\rho\sigma}
-i\hat{\Omega}_{\mu}{}^{\rho}\hat{\varphi}_{\rho\nu}
+i\hat{\Omega}_{\nu}{}^{\rho}\hat{\varphi}_{\rho\mu}\bigr)
\notag\\
& \quad - \frac{i}{2}
\bigl([\hat{\varphi}_{\mu\rho},\hat{\varphi}_{\nu}{}^{\rho}]
+i\hat{\Omega}^{\rho,}{}_{\mu\sigma}D_{\rho}\hat{\varphi}_{\nu}{}^{\sigma}
-i\hat{\Omega}^{\rho,}{}_{\nu\sigma}D_{\rho}\hat{\varphi}_{\mu}{}^{\sigma}
+\hat{\Omega}^{-}_{\mu\nu,\rho\sigma}\hat{\varphi}^{\rho\sigma}
-\hat{\Omega}_{\rho\sigma,}{}^{\rho\sigma}\hat{\varphi}_{\mu\nu}\bigr)
, 
\notag 
\\
\bar{Q} \hat{\Lambda}_{\mu} & =  2 K_{\mu} - \sqrt{2} (D_{\mu} \hat{\varphi} - F_{\mu \nu} \hat{\Omega}^{\nu}) - 2 \sqrt{2} (D^{\nu} \hat{\varphi}_{\mu \nu} - F^{\nu \rho} \hat{\Omega}_{\rho, \mu \nu}).
\label{eq:VW_modified_transformation1}
\end{align}
As in the half twist case, we find that the supersymmetry transformations of
the auxiliary fields are given by
\begin{align}
\bar{Q}D_{\mu\nu}
&=
(D_{\mu}\Lambda_{\nu}-D_{\nu}\Lambda_{\mu})^{-}
-\frac{i}{2\sqrt{2}}
[\hat{\varphi}_{\mu\rho},\hat{\bar{\Lambda}}_{\nu}{}^{\rho}]
+\frac{i}{2\sqrt{2}}
[\hat{\varphi}_{\nu\rho},\hat{\bar{\Lambda}}_{\mu}{}^{\rho}]
\notag\\
&\qquad{}
+\frac{1}{2\sqrt{2}}
\hat{\Omega}^{\rho,}{}_{\mu\sigma}D_{\rho}\hat{\bar{\Lambda}}_{\nu}{}^{\sigma}
-\frac{1}{2\sqrt{2}}
\hat{\Omega}^{\rho,}{}_{\nu\sigma}D_{\rho}\hat{\bar{\Lambda}}_{\mu}{}^{\sigma}
-\frac{i}{2\sqrt{2}}
\hat{\Omega}^{-}_{\mu\nu,\rho\sigma}\hat{\bar{\Lambda}}^{\rho\sigma}
+\frac{i}{2\sqrt{2}}
\hat{\Omega}_{\rho\sigma,}{}^{\rho\sigma}\hat{\bar{\Lambda}}_{\mu\nu}
\notag\\
&\qquad{}
+\frac{i}{\sqrt{2}}[\hat{\varphi},\hat{\bar{\Lambda}}_{\mu\nu}]
-\frac{i}{\sqrt{2}}\hat{\Omega}^{\rho}D_{\rho}\hat{\bar{\Lambda}}_{\mu\nu}
+\frac{i}{\sqrt{2}}\bigl(
\hat{\Omega}_{\mu}{}^{\rho}\hat{\bar{\Lambda}}_{\rho\nu}
-\hat{\Omega}_{\nu}{}^{\rho}\hat{\bar{\Lambda}}_{\rho\mu}
\bigr)
\notag\\
&\qquad{}
+\sqrt{2}i[\varphi,\bar{\Lambda}_{\mu\nu}]
-\sqrt{2}\Omega^{\rho}D_{\rho}\bar{\Lambda}_{\mu\nu}
+\sqrt{2}\bigl(\Omega_{\mu} {}^{\rho}\bar{\Lambda}_{\rho\nu}
-\Omega_{\nu} {}^{\rho} \bar{\Lambda}_{\rho \mu}\bigr)
\notag\\
&\qquad{}
-\frac{i}{\sqrt{2}}[\hat{\varphi}_{\mu\nu},\hat{\bar{\Lambda}}]
+\frac{1}{\sqrt{2}}\hat{\Omega}^{\rho,}{}_{\mu\nu}D_{\rho}\hat{\bar{\Lambda}},
\notag\\
\bar{Q}K_{\mu}
&=
-D_{\mu}\hat{\bar{\Lambda}}-2D^{\nu}\hat{\bar{\Lambda}}_{\mu\nu}
-\sqrt{2}i[\hat{\varphi}_{\mu\nu},\Lambda^{\nu}]
+\sqrt{2}\hat{\Omega}^{\rho,}{}_{\mu\nu}D_{\rho}\Lambda^{\nu}
+\sqrt{2}\hat{\Omega}^{\rho\nu,}{}_{\mu\nu}\Lambda_{\rho}
\notag\\
&\qquad{}
-\frac{i}{\sqrt{2}}[\hat{\varphi},\Lambda_{\mu}]
-\frac{1}{\sqrt{2}}\hat{\Omega}^{\rho}D_{\rho}\Lambda_{\mu}
+\frac{1}{\sqrt{2}}\hat{\Omega}_{\mu}{}^{\nu}\Lambda_{\nu}
\notag\\
&\qquad{}
+\sqrt{2}i[\varphi,\hat{\Lambda}_{\mu}]
-\sqrt{2}\Omega^{\rho}D_{\rho}\hat{\Lambda}_{\mu}
+\sqrt{2}\Omega_{\mu}{}^{\nu}\hat{\Lambda}_{\nu}.
\label{eq103cc}
\end{align}
We redefine the auxiliary fields as 
\begin{align}
H_{\mu \nu} & = D_{\mu \nu} - F^{-}_{\mu \nu} 
\notag \\
& \quad 
+ \frac{i}{2} \bigl([\hat{\varphi},\hat{\varphi}_{\mu\nu}]
+i\hat{\Omega}^{\rho}D_{\rho}\hat{\varphi}_{\mu\nu}
-i\hat{\Omega}^{\rho,}{}_{\mu\nu}D_{\rho}\hat{\varphi}
+i\hat{\Omega}^{\rho,}{}_{\mu\nu}\hat{\Omega}^{\sigma}F_{\rho\sigma}
-i\hat{\Omega}_{\mu}{}^{\rho}\hat{\varphi}_{\rho\nu}
+i\hat{\Omega}_{\nu}{}^{\rho}\hat{\varphi}_{\rho\mu}\bigr)
\notag\\
& \quad - \frac{i}{4}
\bigl([\hat{\varphi}_{\mu\rho},\hat{\varphi}_{\nu}{}^{\rho}]
+i\hat{\Omega}^{\rho,}{}_{\mu\sigma}D_{\rho}\hat{\varphi}_{\nu}{}^{\sigma}
-i\hat{\Omega}^{\rho,}{}_{\nu\sigma}D_{\rho}\hat{\varphi}_{\mu}{}^{\sigma}
+\hat{\Omega}^{-}_{\mu\nu,\rho\sigma}\hat{\varphi}^{\rho\sigma}
-\hat{\Omega}_{\rho\sigma,}{}^{\rho\sigma}\hat{\varphi}_{\mu\nu}\bigr)
, \notag \\
G_{\mu} & = K_{\mu} - \frac{1}{\sqrt{2}} (D_{\mu} \hat{\varphi} - F_{\mu \nu} \hat{\Omega}^{\nu}) - \sqrt{2} (D^{\nu} \hat{\varphi}_{\mu \nu} - F^{\nu \rho} \hat{\Omega}_{\rho, \mu \nu}),
\end{align}
so that \eqref{eq103cc} is rewritten in a simple form as 
\begin{align}
\bar{Q} \bar{\Lambda}_{\mu \nu} 
& 
= 2 H_{\mu \nu},
&
\bar{Q} \hat{\Lambda}_{\mu}
&
= 2 G_{\mu}.
\label{eq:VW_aux}
\end{align}
Then the transformations of $H_{\mu \nu}$ and $G_{\mu}$ become
\begin{align}
\bar{Q} H_{\mu \nu} &= \sqrt{2} i [\varphi, \bar{\Lambda}_{\mu
  \nu}] - \sqrt{2} \Omega^{\rho} D_{\rho} \bar{\Lambda}_{\mu \nu} +
  \sqrt{2} (\Omega_{\mu} {}^{\rho} \bar{\Lambda}_{\rho \nu} -
  \Omega_{\nu} {}^{\rho} \bar{\Lambda}_{\rho \mu}), 
\notag 
\\
 \bar{Q} G_{\mu} &= \sqrt{2} i [\varphi, \hat{\Lambda}_{\mu}]
  - \sqrt{2} \Omega^{\rho} D_{\rho} \hat{\Lambda}_{\mu} + \sqrt{2}
  \Omega_{\mu} {}^{\nu} \hat{\Lambda}_{\nu}.
\label{eq:VW_auxiliary_transformation1}
\end{align}
Using the transformations \eqref{eq:VW_aux},
\eqref{eq:VW_auxiliary_transformation1}, 
and \eqref{eq:VW_OS1} for the other fields, 
we find that the action 
\eqref{eq:VW_action} is written in the $\bar{Q}$-exact form up to
the topological term:
\begin{align}
S_2 &=
\bar{Q} \Xi_2
+\int\!d^{4}x\,
\frac{1}{\kappa g^{2}}\textrm{Tr}\biggl[
\frac{1}{4}F_{\mu\nu}\tilde{F}^{\mu\nu}
\biggr],
\notag\\[2mm]
\Xi_2
&=
\int\!d^{4}x\,
\frac{1}{\kappa g^{2}}\textrm{Tr}\biggl[
-\frac{1}{2}F_{\mu\nu}^{-}\bar{\Lambda}^{\mu\nu}
-\frac{1}{4}H_{\mu\nu}\bar{\Lambda}^{\mu\nu}
-\frac{1}{2\sqrt{2}}\Lambda^{\mu}
(D_{\mu}\bar{\varphi}-F_{\mu\nu}\bar{\Omega}^{\nu})
\notag\\
&\qquad\qquad\qquad\quad{}
+\frac{i}{4}\bar{\Lambda}
([\varphi,\bar{\varphi}]+i\Omega^{\mu}D_{\mu}\bar{\varphi}
-i\bar{\Omega}^{\mu}D_{\mu}\varphi+i\bar{\Omega}^{\mu}\Omega^{\nu}F_{\mu\nu})
\notag\\
&\qquad\qquad\qquad\quad{}
-\frac{1}{4}G_{\mu}\hat{\Lambda}^{\mu}
-\frac{1}{2\sqrt{2}}\hat{\Lambda}^{\mu}
(D_{\mu}\hat{\varphi}-F_{\mu\nu}\hat{\Omega}^{\nu})
+\frac{1}{\sqrt{2}}\hat{\Lambda}_{\nu}
(D_{\mu}\hat{\varphi}^{\mu\nu}-F_{\mu\rho}\hat{\Omega}^{\rho,\mu\nu})
\notag\\
&\qquad\qquad\qquad\quad{}
+\frac{i}{4}\hat{\bar{\Lambda}}
([\bar{\varphi},\hat{\varphi}]+i\bar{\Omega}^{\mu}D_{\mu}\hat{\varphi}
-i\hat{\Omega}^{\mu}D_{\mu}\bar{\varphi}
+i\hat{\Omega}^{\mu}\bar{\Omega}^{\nu}F_{\mu\nu})
\notag\\
&\qquad\qquad\qquad\quad{}
-\frac{i}{4}\hat{\bar{\Lambda}}^{\mu\nu}
([\bar{\varphi},\hat{\varphi}_{\mu\nu}]
+i\bar{\Omega}^{\rho}D_{\rho}\hat{\varphi}_{\mu\nu}
-i\hat{\Omega}^{\rho,}{}_{\mu\nu}D_{\rho}\bar{\varphi}
+i\hat{\Omega}^{\rho,}{}_{\mu\nu}\bar{\Omega}^{\sigma}F_{\rho\sigma}
-i\bar{\Omega}_{\mu}{}^{\rho}\hat{\varphi}_{\rho\nu}
+i\bar{\Omega}_{\nu}{}^{\rho}\hat{\varphi}_{\rho\mu})
\notag\\
&\qquad\qquad\qquad\quad{}
+\frac{i}{4}\bar{\Lambda}^{\mu\nu}
([\hat{\varphi},\hat{\varphi}_{\mu\nu}]
+i\hat{\Omega}^{\rho}D_{\rho}\hat{\varphi}_{\mu\nu}
-i\hat{\Omega}^{\rho,}{}_{\mu\nu}D_{\rho}\hat{\varphi}
+i\hat{\Omega}^{\rho,}{}_{\mu\nu}\hat{\Omega}^{\sigma}F_{\rho\sigma}
-i\hat{\Omega}_{\mu}{}^{\rho}\hat{\varphi}_{\rho\nu}
+i\hat{\Omega}_{\nu}{}^{\rho}\hat{\varphi}_{\rho\mu})
\notag\\
&\qquad\qquad\qquad\quad{}
-\frac{i}{8}\bar{\Lambda}^{\mu\nu}
([\hat{\varphi}_{\mu\rho},\hat{\varphi}_{\nu}{}^{\rho}]
+i\hat{\Omega}^{\rho,}{}_{\mu\sigma}D_{\rho}\hat{\varphi}_{\nu}{}^{\sigma}
-i\hat{\Omega}^{\rho,}{}_{\nu\sigma}D_{\rho}\hat{\varphi}_{\mu}{}^{\sigma}
+\hat{\Omega}_{\mu\nu,\rho\sigma}\hat{\varphi}^{\rho\sigma}
-\hat{\Omega}_{\rho\sigma,}{}^{\rho\sigma}\hat{\varphi}_{\mu\nu})
\biggr].
\label{eq602}
\end{align}
We next study the off-shell transformations generated by $\hat{\bar{Q}}$.
We modify the $\hat{\bar{Q}}$-transformations of 
$\hat{\bar{\Lambda}}_{\mu \nu}$, $\Lambda_{\mu}$ as
\begin{align}
\hat{\bar{Q}} \hat{\bar{\Lambda}}_{\mu \nu} & = 
2 D_{\mu \nu} - 2 F^{-}_{\mu \nu} 
\notag\\
& \quad
- i \bigl([\hat{\varphi},\hat{\varphi}_{\mu\nu}]
+i\hat{\Omega}^{\rho}D_{\rho}\hat{\varphi}_{\mu\nu}
-i\hat{\Omega}^{\rho,}{}_{\mu\nu}D_{\rho}\hat{\varphi}
+i\hat{\Omega}^{\rho,}{}_{\mu\nu}\hat{\Omega}^{\sigma}F_{\rho\sigma}
-i\hat{\Omega}_{\mu}{}^{\rho}\hat{\varphi}_{\rho\nu}
+i\hat{\Omega}_{\nu}{}^{\rho}\hat{\varphi}_{\rho\mu}\bigr)
\notag\\
& \quad - \frac{i}{2}
\bigl([\hat{\varphi}_{\mu\rho},\hat{\varphi}_{\nu}{}^{\rho}]
+i\hat{\Omega}^{\rho,}{}_{\mu\sigma}D_{\rho}\hat{\varphi}_{\nu}{}^{\sigma}
-i\hat{\Omega}^{\rho,}{}_{\nu\sigma}D_{\rho}\hat{\varphi}_{\mu}{}^{\sigma}
+\hat{\Omega}^{-}_{\mu\nu,\rho\sigma}\hat{\varphi}^{\rho\sigma}
-\hat{\Omega}_{\rho\sigma,}{}^{\rho\sigma}\hat{\varphi}_{\mu\nu}\bigr)
, 
\notag \\
\hat{\bar{Q}} \Lambda_{\mu}
& = 
- 2 K_{\mu} 
- \sqrt{2} 
(D_{\mu} \hat{\varphi} - F_{\mu \nu} \hat{\Omega}^{\nu}) 
+ 2 \sqrt{2}
(D^{\nu} \hat{\varphi}_{\mu \nu} - F^{\nu \rho} \hat{\Omega}_{\rho, \mu \nu}).
\label{eq:VW_modified_transformation2}
\end{align}
The $\hat{\bar{Q}}$-transformations of the auxiliary fields 
are given by 
\begin{align}
\hat{\bar{Q}}D_{\mu\nu}
&=
(D_{\mu}\hat{\Lambda}_{\nu}-D_{\nu}\hat{\Lambda}_{\mu})^{-}
+\frac{i}{2\sqrt{2}}
[\hat{\varphi}_{\mu\rho},\bar{\Lambda}_{\nu}{}^{\rho}]
-\frac{i}{2\sqrt{2}}
[\hat{\varphi}_{\nu\rho},\bar{\Lambda}_{\mu}{}^{\rho}]
\notag\\
&\qquad{}
-\frac{1}{2\sqrt{2}}
\hat{\Omega}^{\rho,}{}_{\mu\sigma}D_{\rho}\bar{\Lambda}_{\nu}{}^{\sigma}
+\frac{1}{2\sqrt{2}}
\hat{\Omega}^{\rho,}{}_{\nu\sigma}D_{\rho}\bar{\Lambda}_{\mu}{}^{\sigma}
+\frac{i}{2\sqrt{2}}
\hat{\Omega}^{-}_{\mu\nu,\rho\sigma}\bar{\Lambda}^{\rho\sigma}
-\frac{i}{2\sqrt{2}}
\hat{\Omega}_{\rho\sigma,}{}^{\rho\sigma}\bar{\Lambda}_{\mu\nu}
\notag\\
&\qquad{}
+\frac{i}{\sqrt{2}}[\hat{\varphi},\bar{\Lambda}_{\mu\nu}]
-\frac{i}{\sqrt{2}}\hat{\Omega}^{\rho}D_{\rho}\bar{\Lambda}_{\mu\nu}
+\frac{i}{\sqrt{2}}\bigl(
\hat{\Omega}_{\mu}{}^{\rho}\bar{\Lambda}_{\rho\nu}
-\hat{\Omega}_{\nu}{}^{\rho}\bar{\Lambda}_{\rho\mu}
\bigr)
\notag\\
&\qquad{}
-\sqrt{2}i[\bar{\varphi},\hat{\bar{\Lambda}}_{\mu\nu}]
+\sqrt{2}\bar{\Omega}^{\rho}D_{\rho}\hat{\bar{\Lambda}}_{\mu\nu}
-\sqrt{2}\bigl(\bar{\Omega}_{\mu} {}^{\rho}\hat{\bar{\Lambda}}_{\rho\nu}
-\bar{\Omega}_{\nu} {}^{\rho}\hat{\bar{\Lambda}}_{\rho\mu}\bigr)
\notag\\
&\qquad{}
+\frac{i}{\sqrt{2}}[\hat{\varphi}_{\mu\nu},\bar{\Lambda}]
-\frac{1}{\sqrt{2}}\hat{\Omega}^{\rho,}{}_{\mu\nu}D_{\rho}\bar{\Lambda},
\notag\\[2mm]
\hat{\bar{Q}}K_{\mu}
&=
D_{\mu}\bar{\Lambda}+2D^{\nu}\bar{\Lambda}_{\mu\nu}
-\sqrt{2}i[\hat{\varphi}_{\mu\nu},\hat{\Lambda}^{\nu}]
+\sqrt{2}\hat{\Omega}^{\rho,}{}_{\mu\nu}D_{\rho}\hat{\Lambda}^{\nu}
+\sqrt{2}\hat{\Omega}^{\rho\nu,}{}_{\mu\nu}\hat{\Lambda}_{\rho}
\notag\\
&\qquad{}
+\frac{i}{\sqrt{2}}[\hat{\varphi},\hat{\Lambda}_{\mu}]
-\frac{1}{\sqrt{2}}\hat{\Omega}^{\nu}D_{\nu}\hat{\Lambda}_{\mu}
+\frac{1}{\sqrt{2}}\hat{\Omega}_{\mu}{}^{\nu}\hat{\Lambda}_{\nu}
\notag\\
&\qquad{}
+\sqrt{2}i[\bar{\varphi},\Lambda_{\mu}]
-\sqrt{2}\bar{\Omega}^{\nu}D_{\nu}\Lambda_{\mu}
+\sqrt{2}\bar{\Omega}_{\mu}{}^{\nu}\Lambda_{\nu}.
\label{eq205}
\end{align}
We redefine the auxiliary fields as 
\begin{align}
\hat{H}_{\mu\nu}&=
D_{\mu\nu}-F^{-}_{\mu\nu} 
\notag\\
&\quad{}
-\frac{i}{2}\bigl([\hat{\varphi},\hat{\varphi}_{\mu\nu}]
+i\hat{\Omega}^{\rho}D_{\rho}\hat{\varphi}_{\mu\nu}
-i\hat{\Omega}^{\rho,}{}_{\mu\nu}D_{\rho}\hat{\varphi}
+i\hat{\Omega}^{\rho,}{}_{\mu\nu}\hat{\Omega}^{\sigma}F_{\rho\sigma}
-i\hat{\Omega}_{\mu}{}^{\rho}\hat{\varphi}_{\rho\nu}
+i\hat{\Omega}_{\nu}{}^{\rho}\hat{\varphi}_{\rho\mu}\bigr)
\notag\\
&\quad{}
-\frac{i}{4}
\bigl([\hat{\varphi}_{\mu\rho},\hat{\varphi}_{\nu}{}^{\rho}]
+i\hat{\Omega}^{\rho,}{}_{\mu\sigma}D_{\rho}\hat{\varphi}_{\nu}{}^{\sigma}
-i\hat{\Omega}^{\rho,}{}_{\nu\sigma}D_{\rho}\hat{\varphi}_{\mu}{}^{\sigma}
+\hat{\Omega}^{-}_{\mu\nu,\rho\sigma}\hat{\varphi}^{\rho\sigma}
-\hat{\Omega}_{\rho\sigma,}{}^{\rho\sigma}\hat{\varphi}_{\mu\nu}\bigr), 
\notag \\
\hat{G}_{\mu}&=
K_{\mu}+\frac{1}{\sqrt{2}}
(D_{\mu}\hat{\varphi}-F_{\mu\nu}\hat{\Omega}^{\nu}) 
-\sqrt{2}
(D^{\nu}\hat{\varphi}_{\mu\nu}-F^{\nu\rho}\hat{\Omega}_{\rho,\mu\nu}),
\end{align}
such that \eqref{eq205} is simply rewritten as 
\begin{align}
\hat{\bar{Q}} \hat{\bar{\Lambda}}_{\mu \nu}
&
= 2 \hat{H}_{\mu \nu},
&
\hat{\bar{Q}} \Lambda_{\mu}
&
= - 2 \hat{G}_{\mu}.
\end{align}
Then the transformations for $\hat{H}_{\mu\nu}$ and $\hat{G}_{\mu}$ are 
\begin{align}
 \hat{\bar{Q}} \hat{H}_{\mu\nu}
 &=
 -\sqrt{2}i[\bar{\varphi},\hat{\bar{\Lambda}}_{\mu\nu}]
 +\sqrt{2}\bar{\Omega}^{\rho}D_{\rho}\hat{\bar{\Lambda}}_{\mu\nu}
 -\sqrt{2}(\bar{\Omega}_{\mu}{}^{\rho}\hat{\bar{\Lambda}}_{\rho\nu}
 -\bar{\Omega}_{\nu}{}^{\rho}\hat{\bar{\Lambda}}_{\rho\mu})
, 
\notag 
\\
 \hat{\bar{Q}} \hat{G}_{\mu}
 &=
 \sqrt{2}i[\bar{\varphi},\Lambda_{\mu}]
 -\sqrt{2}\bar{\Omega}^{\nu}D_{\nu}\Lambda_{\mu}
 +\sqrt{2}\bar{\Omega}_{\mu}{}^{\nu}\Lambda_{\nu}.
\label{eq:VW_auxiliary_transformation2}
\end{align}
Using the transformations \eqref{eq:VW_modified_transformation2}, 
\eqref{eq:VW_auxiliary_transformation2} 
and \eqref{eq:VW_OS2} for the other fields, 
we find that the action is written in the $\hat{\bar{Q}}$-exact form
up to the topological term:
\begin{align}
S_2 &=
\hat{\bar{Q}} \Xi_2'
+\int\!d^{4}x\,
\frac{1}{\kappa g^{2}}\textrm{Tr}\biggl[
\frac{1}{4}F_{\mu\nu}\tilde{F}^{\mu\nu}
\biggr],
\notag\\
\Xi_2'
&=
\int\!d^{4}x\,
\frac{1}{\kappa g^{2}}\textrm{Tr}\biggl[
-\frac{1}{2}F_{\mu\nu}^{-}\hat{\bar{\Lambda}}^{\mu\nu}
-\frac{1}{4}\hat{H}_{\mu\nu}\hat{\bar{\Lambda}}^{\mu\nu}
+\frac{1}{2\sqrt{2}}\hat{\Lambda}^{\mu}
(D_{\mu}\varphi-F_{\mu\nu}\Omega^{\nu})
\notag\\
&\qquad\qquad\qquad\quad{}
-\frac{i}{4}\hat{\bar{\Lambda}}
([\varphi,\bar{\varphi}]+i\Omega^{\mu}D_{\mu}\bar{\varphi}
-i\bar{\Omega}^{\mu}D_{\mu}\varphi+i\bar{\Omega}^{\mu}\Omega^{\nu}F_{\mu\nu})
\notag\\
&\qquad\qquad\qquad\quad{}
-\frac{1}{4}\hat{G}_{\mu}\Lambda^{\mu}
+\frac{1}{2\sqrt{2}}\Lambda^{\mu}
(D_{\mu}\hat{\varphi}-F_{\mu\nu}\hat{\Omega}^{\nu})
-\frac{1}{\sqrt{2}}\Lambda_{\nu}
(D_{\mu}\hat{\varphi}^{\mu\nu}-F_{\mu\rho}\hat{\Omega}^{\rho,\mu\nu})
\notag\\
&\qquad\qquad\qquad\quad{}
+\frac{i}{4}\bar{\Lambda}
([\varphi,\hat{\varphi}]+i\Omega^{\mu}D_{\mu}\hat{\varphi}
-i\hat{\Omega}^{\mu}D_{\mu}\varphi
+i\hat{\Omega}^{\mu}\Omega^{\nu}F_{\mu\nu})
\notag\\
&\qquad\qquad\qquad\quad{}
-\frac{i}{4}\bar{\Lambda}^{\mu\nu}
([\varphi,\hat{\varphi}_{\mu\nu}]
+i\Omega^{\rho}D_{\rho}\hat{\varphi}_{\mu\nu}
-i\hat{\Omega}^{\rho,}{}_{\mu\nu}D_{\rho}\varphi
+i\hat{\Omega}^{\rho,}{}_{\mu\nu}\Omega^{\sigma}F_{\rho\sigma}
-i\Omega_{\mu}{}^{\rho}\hat{\varphi}_{\rho\nu}
+i\Omega_{\nu}{}^{\rho}\hat{\varphi}_{\rho\mu})
\notag\\
&\qquad\qquad\qquad\quad{}
-\frac{i}{4}\hat{\bar{\Lambda}}^{\mu\nu}
([\hat{\varphi},\hat{\varphi}_{\mu\nu}]
+i\hat{\Omega}^{\rho}D_{\rho}\hat{\varphi}_{\mu\nu}
-i\hat{\Omega}^{\rho,}{}_{\mu\nu}D_{\rho}\hat{\varphi}
+i\hat{\Omega}^{\rho,}{}_{\mu\nu}\hat{\Omega}^{\sigma}F_{\rho\sigma}
-i\hat{\Omega}_{\mu}{}^{\rho}\hat{\varphi}_{\rho\nu}
+i\hat{\Omega}_{\nu}{}^{\rho}\hat{\varphi}_{\rho\mu})
\notag\\
&\qquad\qquad\qquad\quad{}
+\frac{i}{8}\hat{\bar{\Lambda}}^{\mu\nu}
([\hat{\varphi}_{\mu\rho},\hat{\varphi}_{\nu}{}^{\rho}]
+i\hat{\Omega}^{\rho,}{}_{\mu\sigma}D_{\rho}\hat{\varphi}_{\nu}{}^{\sigma}
-i\hat{\Omega}^{\rho,}{}_{\nu\sigma}D_{\rho}\hat{\varphi}_{\mu}{}^{\sigma}
+\hat{\Omega}_{\mu\nu,\rho\sigma}\hat{\varphi}^{\rho\sigma}
-\hat{\Omega}_{\rho\sigma,}{}^{\rho\sigma}\hat{\varphi}_{\mu\nu})
\biggr].
\label{eq606}
\end{align}
The transformations of a field $\Psi$ by $\bar{Q}$ and $\hat{\bar{Q}}$ 
satisfy the following off-shell algebra:
\begin{align}
& \bar{Q}^2 \Psi
= 2 \sqrt{2} 
\bigl(\,\delta_{\mathrm{gauge}} (\varphi) 
+ 
\delta_{\mathrm{Lorentz}} (\Omega) 
\,\bigr)
\Psi, 
\notag \\
& \hat{\bar{Q}}^2 \Psi
= - 2 \sqrt{2} \bigl(\,\delta_{\mathrm{gauge}} (\bar{\varphi}) 
+ \delta_{\mathrm{Lorentz}} (\bar{\Omega}) 
\,\bigr)
\Psi, 
\notag \\
& \{\bar{Q}, \hat{\bar{Q}} \} \Psi
= 2 \sqrt{2} \bigl(\,\delta_{\mathrm{gauge}} (\hat{\varphi}) 
+ 
\delta_{\mathrm{Lorentz}} (\hat{\Omega}) 
\,\bigr)
\Psi.
\end{align}

In the undeformed case, the action is written in the exact form 
by the two scalar supercharges simultaneously \cite{Yamron,
Vafa:1994tf, Dijkgraaf:1996tz}. 
We find that this is also true in the deformed theory.
The action is expressed as 
\begin{align}
S_2 = \bar{Q} \hat{\bar{Q}} \mathcal{F}
{}
+\int\!d^{4}x\,
\frac{1}{\kappa g^{2}}\textrm{Tr}\biggl[
\frac{1}{4}F_{\mu\nu}\tilde{F}^{\mu\nu}
\biggr],
\label{eq:balanced}
\end{align}
where
\begin{align}
\mathcal{F}
&=
\int \! d^4 x \ 
\frac{1}{\kappa
 g^2} 
\mathrm{Tr}
\biggl[
- \frac{1}{2 \sqrt{2}} \hat{\varphi}^{\mu \nu} F_{\mu \nu}^{-}
+ \frac{1}{8} \bar{\Lambda}^{\mu \nu} \hat{\bar{\Lambda}}_{\mu \nu} +
 \frac{1}{8} \Lambda^{\mu} \Lambda_{\mu} - \frac{1}{8} \bar{\Lambda}
 \hat{\bar{\Lambda}} + \frac{i}{24 \sqrt{2}} \hat{\varphi}^{\mu \nu}
 [\hat{\varphi}_{\mu} {}^{\lambda}, \hat{\varphi}_{\lambda \nu}]
\notag\\
&\qquad\qquad\qquad\quad{}
+\frac{1}{16\sqrt{2}}\hat{\varphi}^{\mu\nu}
(\hat{\Omega}^{\rho,}{}_{\mu\sigma}D_{\rho}\hat{\varphi}_{\nu}{}^{\sigma}
-\hat{\Omega}^{\rho,}{}_{\nu\sigma}D_{\rho}\hat{\varphi}_{\mu}{}^{\sigma}
-i\hat{\Omega}_{\mu\nu,\rho\sigma}\hat{\varphi}^{\rho\sigma}
+i\hat{\Omega}_{\rho\sigma,}{}^{\rho\sigma}\hat{\varphi}_{\mu\nu})
\notag\\
&\qquad\qquad\qquad\quad{}
+\frac{3}{2\sqrt{2}}\hat{\Omega}^{[\rho,\mu\nu]}
\biggl(A_{[\mu}F_{\nu\rho]}-\frac{i}{3}A_{[\mu}A_{\nu}A_{\rho]}\biggr)
\biggr].
\end{align}
Here the three indices in the square bracket are totally antisymmetrized 
with the normalization $1/3!$.
Note that $\mathcal{F}$ is gauge invariant.
This is because the gauge transformation 
of $\mathcal{F}$ by the gauge parameter $\alpha$ is computed as 
\begin{gather}
\delta_{\mathrm{gauge}}(\alpha)\mathcal{F}
=
\int \! d^4 x \ 
\frac{1}{\kappa g^2}\mathrm{Tr}\biggl[
-\frac{1}{2\sqrt{2}}
(\hat{\Omega}_{\rho}{}^{\mu,\nu\rho}-\hat{\Omega}_{\rho}{}^{\nu,\mu\rho})
F_{\mu\nu}\alpha
\biggr],
\label{eq:VW_gauge_tr_F}
\end{gather}
and $\hat{\Omega}_{\rho}{}^{\mu,\nu\rho}$ 
is symmetric with respect to $\mu$ and $\nu$ from \eqref{eq:VW_omega_condition2}.

\subsection{The Marcus twist}
In the case of the Marcus twist, 
there are two scalar supercharges $Q$ and $\bar{Q}$, which have the opposite chirality. 
As studied in \cite{Marcus:1995mq, Kapustin:2006pk}, in the undeformed case, 
one cannot make both $Q$
and $\bar{Q}$ off-shell but can make only their linear combination off-shell.
This charge plays an important role for studying the generalized Langlands
duality of ${\cal N}=4$ theory compactified on a Riemann surface 
\cite{Kapustin:2006pk}. 
Now we will examine whether this off-shell supersymmetry structure is
kept under the $\Omega$-deformation.

We first study the off-shell supersymmetry generated by $\bar{Q}$.
In order to construct off-shell supersymmetry, we introduce the
auxiliary fields $K$, $K_{\mu \nu}$, $D_{\mu \nu}$.
Then we add the quadratic terms of the auxiliary fields to 
the action \eqref{eq:on-shell_action} as  
\begin{align}
S_3 =  S 
+ \int \! d^4 x \ \frac{1}{\kappa g^2} 
 \mathrm{Tr}
\left[
- \frac{1}{2} (D_{\mu \nu})^2 - \frac{1}{2} (K_{\mu \nu})^2 -
 \frac{1}{2} K^2
\right].
\label{eq:MOS_action}
\end{align}
We modify the transformations of the fields $\Lambda$, $\Lambda_{\mu
\nu}$, $\bar{\Lambda}_{\mu \nu}$ as 
\begin{align}
\bar{Q} \Lambda =& 2 K + \sqrt{2} D_{\mu} \varphi^{\mu}, 
\notag 
\\
\bar{Q} \Lambda_{\mu \nu} =& 2 K_{\mu \nu} + \sqrt{2} (D_{\mu}
 \varphi_{\nu} - D_{\nu} \varphi_{\mu})^{+}, 
\notag 
\\
\bar{Q} \bar{\Lambda}_{\mu \nu} =& 2 D_{\mu \nu} - 2 F^{-}_{\mu \nu} + i
 [\varphi_{\mu}, \varphi_{\nu}]^{-}.
\label{eq:Marcus_modified_transformation1}
\end{align}
The transformations of the auxiliary fields are determined 
as 
\begin{align}
\bar{Q}K
&=
D_{\mu}\bar{\Lambda}^{\mu}
-\frac{i}{\sqrt{2}}[\Lambda_{\mu},\varphi^{\mu}]
+\sqrt{2}i([\varphi,\Lambda]+i\Omega^{\mu}D_{\mu}\Lambda),
\notag\\
\bar{Q}K_{\mu\nu}
&=
(D_{\mu}\bar{\Lambda}_{\nu}-D_{\nu}\bar{\Lambda}_{\mu})^{+}
+\frac{i}{\sqrt{2}}
([\varphi_{\mu},\Lambda_{\nu}]-[\varphi_{\nu},\Lambda_{\mu}])^{+}
\notag\\
&\quad{}
+\sqrt{2}i[\varphi,\Lambda_{\mu\nu}]
-\sqrt{2}\Omega^{\lambda}D_{\lambda}\Lambda_{\mu\nu}
+\sqrt{2}(\Omega_{\mu}{}^{\lambda}\Lambda_{\lambda\nu}
-\Omega_{\nu}{}^{\lambda}\Lambda_{\lambda\mu}),
\notag\\
\bar{Q}D_{\mu\nu}
&=
(D_{\mu}\Lambda_{\nu}-D_{\nu}\Lambda_{\mu})^{-}
-\frac{i}{\sqrt{2}}
([\varphi_{\mu},\bar{\Lambda}_{\nu}]-[\varphi_{\nu},\bar{\Lambda}_{\mu}])^{-}
\notag\\
&\quad{}
+\sqrt{2}i[\varphi,\bar{\Lambda}_{\mu\nu}]
-\sqrt{2}\Omega^{\lambda}D_{\lambda}\bar{\Lambda}_{\mu\nu}
+\sqrt{2}(\Omega_{\mu}{}^{\lambda}\bar{\Lambda}_{\lambda\nu}
-\Omega_{\nu}{}^{\lambda}\bar{\Lambda}_{\lambda\mu}).
\label{eq:Marcus_auxiliary_transformation1}
\end{align}
We redefine the auxiliary fields as 
\begin{align}
G = & K + \frac{1}{\sqrt{2}} D_{\mu} \varphi^{\mu}, 
\notag \\
G_{\mu \nu} = & K_{\mu \nu} + \frac{1}{\sqrt{2}} (D_{\mu} \varphi_{\nu}
 - D_{\nu} \varphi_{\mu})^{+},
\notag \\
H_{\mu \nu} = & D_{\mu \nu} - F^{-}_{\mu \nu} + \frac{i}{2}
 [\varphi_{\mu}, \varphi_{\nu}]^{-}, 
\label{eq:Marcus_redefinition1}
\end{align}
such that \eqref{eq:Marcus_auxiliary_transformation1}
 takes a simple form as
\begin{align}
\bar{Q}\Lambda&=2G, &
\bar{Q}\Lambda_{\mu\nu}&=2G_{\mu\nu}, &
\bar{Q}\bar{\Lambda}_{\mu\nu}&=2H_{\mu\nu}.
\label{eq:MT_aux1}
\end{align}
Then the transformations for $G$, $G_{\mu\nu}$ and $H_{\mu\nu}$ are 
given by 
\begin{align}
\bar{Q} G
&=
\sqrt{2}i([\varphi,\Lambda]+i\Omega^{\mu}D_{\mu}\Lambda), 
\notag 
\\
\bar{Q} G_{\mu\nu}
&=
\sqrt{2}i[\varphi,\Lambda_{\mu\nu}]
-\sqrt{2}\Omega^{\lambda}D_{\lambda}\Lambda_{\mu\nu}
+\sqrt{2}(\Omega_{\mu}{}^{\lambda}\Lambda_{\lambda\nu}
-\Omega_{\nu}{}^{\lambda}\Lambda_{\lambda\mu}), 
\notag 
\\
\bar{Q} H_{\mu\nu}
&=
 \sqrt{2}i[\varphi,\bar{\Lambda}_{\mu\nu}]
 -\sqrt{2}\Omega^{\lambda}D_{\lambda}\bar{\Lambda}_{\mu\nu}
 +\sqrt{2}(\Omega_{\mu}{}^{\lambda}\bar{\Lambda}_{\lambda\nu}
 -\Omega_{\nu}{}^{\lambda}\bar{\Lambda}_{\lambda\mu}).
\label{eq:MT_aux2}
\end{align}
Using the transformations,  
\eqref{eq:MT_aux1}, 
\eqref{eq:MT_aux2}
and \eqref{eq:Marcus_OS1} for the other fields,  
we find that the action \eqref{eq:MOS_action} 
is written in the $\bar{Q}$-exact form up to the 
topological term:
\begin{align}
S_3 & =
\bar{Q} \Xi_3
+ \int\!d^{4}x\,
\frac{1}{\kappa g^{2}}\textrm{Tr}\biggl[
\frac{1}{4}F_{\mu\nu}\tilde{F}^{\mu\nu}
\biggr], 
\end{align}
where 
\begin{align}
\Xi_3
&=
\int\!d^{4}x\,
\frac{1}{\kappa g^{2}}\textrm{Tr}\biggl[
-\frac{1}{2}F_{\mu\nu}^{-}\bar{\Lambda}^{\mu\nu}
+\frac{i}{4}\bar{\Lambda}^{\mu\nu}[\varphi_{\mu},\varphi_{\nu}]^{-}
-\frac{1}{4}H_{\mu\nu}\bar{\Lambda}^{\mu\nu}
\notag\\
&\qquad\qquad\qquad\qquad\quad{}
-\frac{1}{4}\Lambda_{\mu\nu}G^{\mu\nu}
+\frac{1}{2\sqrt{2}}\Lambda^{\mu\nu}
(D_{\mu}\varphi_{\nu}-D_{\nu}\varphi_{\mu})^{+}
\notag \\
& \qquad\qquad\qquad\qquad\quad{}
-\frac{1}{4}\Lambda G
+\frac{i}{4}\bar{\Lambda}\bigl([\varphi,\bar{\varphi}]
+i\Omega^{\mu}D_{\mu}\bar{\varphi}-i\bar{\Omega}^{\mu}D_{\mu}\varphi
+i\bar{\Omega}^{\mu}\Omega^{\nu}F_{\mu\nu}\bigr)
\notag\\
&\qquad\qquad\qquad\qquad\quad{}
+\frac{1}{2\sqrt{2}}\Lambda D_{\mu}\varphi^{\mu}
-\frac{1}{2\sqrt{2}}\Lambda^{\mu}
(D_{\mu}\bar{\varphi}-F_{\mu\nu}\bar{\Omega}^{\nu})
\notag\\
&\qquad\qquad\qquad\qquad\quad{}
-\frac{i}{4}\bar{\Lambda}^{\mu}\bigl([\bar{\varphi},\varphi_{\mu}]
+i\bar{\Omega}^{\nu}D_{\nu}\varphi_{\mu}
-i\bar{\Omega}_{\mu}{}^{\nu}\varphi_{\nu}\bigr)
\biggr].
\end{align}

We next study the transformations generated by $Q$. 
We modify the transformations of the fields $\bar{\Lambda}$, $\Lambda_{\mu \nu}$
and $\bar{\Lambda}_{\mu \nu}$ as 
\begin{align}
Q \bar{\Lambda} =& 2 K + \sqrt{2} D_{\mu} \varphi^{\mu}, 
\notag 
\\
Q \Lambda_{\mu \nu} =& 2 K_{\mu \nu} - 2 F^{+}_{\mu \nu} + i
 [\varphi_{\mu}, \varphi_{\nu}]^{+}, 
\notag 
\\
Q \bar{\Lambda}_{\mu \nu} =& 2 D_{\mu \nu} - \sqrt{2} (D_{\mu}
 \varphi_{\nu} - D_{\nu} \varphi_{\mu})^{-}.
\label{eq:Marcus_modified_transformation2}
\end{align}
We find that the transformations of the auxiliary fields are 
\begin{align}
QK
&=
-D_{\mu}\Lambda^{\mu}
-\frac{i}{\sqrt{2}}[\bar{\Lambda}_{\mu},\varphi^{\mu}]
+\sqrt{2}i([\varphi,\bar{\Lambda}]+i\Omega^{\mu}D_{\mu}\bar{\Lambda}), 
\notag\\
QK_{\mu\nu}
&=
(D_{\mu}\bar{\Lambda}_{\nu}-D_{\nu}\bar{\Lambda}_{\mu})^{+}
+\frac{i}{\sqrt{2}}
([\varphi_{\mu},\Lambda_{\nu}]-[\varphi_{\nu},\Lambda_{\mu}])^{+}
\notag\\
&\quad{}
+\sqrt{2}i[\varphi,\Lambda_{\mu\nu}]
-\sqrt{2}\Omega^{\lambda}D_{\lambda}\Lambda_{\mu\nu}
+\sqrt{2}(\Omega_{\mu}{}^{\lambda}\Lambda_{\lambda\nu}
-\Omega_{\nu}{}^{\lambda}\Lambda_{\lambda\mu}),
\notag\\
QD_{\mu\nu}
&=
(D_{\mu}\Lambda_{\nu}-D_{\nu}\Lambda_{\mu})^{-}
-\frac{i}{\sqrt{2}}
([\varphi_{\mu},\bar{\Lambda}_{\nu}]-[\varphi_{\nu},\bar{\Lambda}_{\mu}])^{-}
\notag\\
&\quad{}
+\sqrt{2}i[\varphi,\bar{\Lambda}_{\mu\nu}]
-\sqrt{2}\Omega^{\lambda}D_{\lambda}\bar{\Lambda}_{\mu\nu}
+\sqrt{2}(\Omega_{\mu}{}^{\lambda}\bar{\Lambda}_{\lambda\nu}
-\Omega_{\nu}{}^{\lambda}\bar{\Lambda}_{\lambda\mu}).
\end{align}
We redefine the auxiliary fields such that 
\eqref{eq:Marcus_modified_transformation2} takes a simple form
as 
\begin{align}
G'_{\mu \nu} & = K_{\mu \nu} - F^{+}_{\mu \nu} + \frac{i}{2}
 [\varphi_{\mu}, \varphi_{\nu}]^{+},
\notag\\
H'_{\mu \nu} & = D_{\mu \nu} - \frac{1}{\sqrt{2}} (D_{\mu} \varphi_{\nu}
 - D_{\nu} \varphi_{\mu})^{-}. 
\end{align}
Then \eqref{eq:Marcus_modified_transformation2} becomes
\begin{align}
Q\bar{\Lambda}&=2G, &
Q\Lambda_{\mu\nu}&=2G'_{\mu\nu}, &
Q\bar{\Lambda}_{\mu\nu}&=2H'_{\mu\nu}.
\label{eq:MT_aux3}
\end{align}
The transformations 
for $G$, $G'_{\mu \nu}$ and $H'_{\mu \nu}$ are 
\begin{align}
Q G =& \sqrt{2}i([\varphi,\bar{\Lambda}] 
+i\Omega^{\mu}D_{\mu}\bar{\Lambda}), 
\notag 
\\
Q G'_{\mu\nu}
=&
\sqrt{2}i[\varphi,\Lambda_{\mu\nu}]
-\sqrt{2}\Omega^{\lambda}D_{\lambda}\Lambda_{\mu\nu}
+\sqrt{2}(\Omega_{\mu}{}^{\lambda}\Lambda_{\lambda\nu}
-\Omega_{\nu}{}^{\lambda}\Lambda_{\lambda\mu}), 
\notag 
\\
Q H'_{\mu\nu}
=&
\sqrt{2}i[\varphi,\bar{\Lambda}_{\mu\nu}]
-\sqrt{2}\Omega^{\lambda}D_{\lambda}\bar{\Lambda}_{\mu\nu}
+\sqrt{2}(\Omega_{\mu}{}^{\lambda}\bar{\Lambda}_{\lambda\nu}
-\Omega_{\nu}{}^{\lambda}\bar{\Lambda}_{\lambda\mu}).
\label{eq:Marcus_auxiliary_transformation2}
\end{align}
Again, we find that the action \eqref{eq:MOS_action} 
is written in the $Q$-exact form up to the topological term:
\begin{align}
S_3 & 
= Q \bar{\Xi}_3
- \int\!d^{4}x\,
\frac{1}{\kappa g^{2}}\textrm{Tr}\biggl[
\frac{1}{4}F_{\mu\nu}\tilde{F}^{\mu\nu}
\biggr], 
\end{align}
where 
\begin{align}
\bar{\Xi}_3
&=
\int\!d^{4}x\,
\frac{1}{\kappa g^{2}}\textrm{Tr}\biggl[
-\frac{1}{2}F_{\mu\nu}^{+}\Lambda^{\mu\nu}
+\frac{i}{4}\Lambda^{\mu\nu}[\varphi_{\mu},\varphi_{\nu}]^{+}
-\frac{1}{4}G'_{\mu\nu}\Lambda^{\mu\nu}
\notag\\[2mm]
&\qquad\qquad\qquad\qquad\quad{}
-\frac{1}{4}\bar{\Lambda}^{\mu\nu}H'_{\mu\nu}
-\frac{1}{2\sqrt{2}}\bar{\Lambda}^{\mu\nu}
(D_{\mu}\varphi_{\nu}-D_{\nu}\varphi_{\mu})^{-}
\notag \\
&\qquad\qquad\qquad\qquad\quad{}
-\frac{1}{4}\bar{\Lambda} G
-\frac{i}{4}\Lambda\bigl([\varphi,\bar{\varphi}]
+i\Omega^{\mu}D_{\mu}\bar{\varphi}-i\bar{\Omega}^{\mu}D_{\mu}\varphi
+i\bar{\Omega}^{\mu}\Omega^{\nu}F_{\mu\nu}\bigr)
\notag\\[2mm]
&\qquad\qquad\qquad\qquad\quad{}
+\frac{1}{2\sqrt{2}}\bar{\Lambda} D_{\mu}\varphi^{\mu}
-\frac{1}{2\sqrt{2}}\bar{\Lambda}^{\mu}
(D_{\mu}\bar{\varphi}-F_{\mu\nu}\bar{\Omega}^{\nu})
\notag\\[2mm]
&\qquad\qquad\qquad\qquad\quad{}
+\frac{i}{4}\bar{\Lambda}^{\mu}\bigl([\bar{\varphi},\varphi_{\mu}]
+i\bar{\Omega}^{\nu}D_{\nu}\varphi_{\mu}
-i\bar{\Omega}_{\mu}{}^{\nu}\varphi_{\nu}\bigr)
\biggr].
\end{align}
The supercharges $Q$ and $\bar{Q}$ satisfy the following on-shell relations on a field $\Psi$
\begin{align}
\bar{Q}^2 \Psi =
Q^2 \Psi
& = 2 \sqrt{2} \bigl(\,\delta_{\mathrm{gauge}} (\varphi)
+ 
\delta_{\mathrm{Lorentz}} (\Omega) 
\,\bigr)
\Psi, 
\label{eq:Marcus_algebra1}
\\
& \{ Q, \bar{Q} \} \Psi = 0.
\label{eq:Marcus_algebra2}
\end{align}
We find that \eqref{eq:Marcus_algebra1} holds off-shell for all the fields 
 but \eqref{eq:Marcus_algebra2} does not hold off-shell on the fields
$\Lambda_{\mu\nu}$, $\bar{\Lambda}_{\mu\nu}$, $K_{\mu \nu}$ and $D_{\mu \nu}$.
Therefore the algebra of symmetry generated by two supercharges $Q$ and $\bar{Q}$ does
not close off-shell.

We can choose the linear combination of the two supercharges 
\begin{align}
\mathcal{Q} = u Q + v \bar{Q}, \quad u,v \in \mathbb{C},
\label{eq:calQ}
\end{align}
such that $\mathcal{Q}$ becomes off-shell.
In the undeformed case, when $u^2+v^2\not=0$, 
the action is shown to be the $\mathcal{Q}$-exact
form up to the topological term \cite{Kapustin:2006pk}.
When $u^2+v^2=0$, the action is not written in the
$\mathcal{Q}$-exact form but it is $\mathcal{Q}$-closed 
\cite{Marcus:1995mq}.
In the following, we show that this property also holds in the deformed
theory.
Since the two supercharges $Q$ and $\bar{Q}$ 
satisfy the relations \eqref{eq:Marcus_algebra1},
\eqref{eq:Marcus_algebra2} on-shell, $\mathcal{Q}$ satisfies the on-shell transformation
\begin{align}
\mathcal{Q}^2 \Psi = 2 \sqrt{2} (u^2 + v^2) 
(\delta_{\mathrm{gauge}} (\varphi) + 
\delta_{\mathrm{Lorentz}} (\Omega))
\Psi. 
\label{eq:Marcus_nilpotent}
\end{align}
In the following, we study the off-shell generalization of the supersymmetry
generated by $\mathcal{Q}$ and examine the $\mathcal{Q}$-exactness of
the action $S_3$ in the cases where $u^2+v^2\not=0$ and $u^2+v^2=0$.
\\
\\
\underline{\textbullet \ $u^2+v^2\not=0$ case}
\\
\\
Since the algebra of $Q$ and $\bar{Q}$ does not close on the fields 
$\Lambda_{\mu \nu}$, $\bar{\Lambda}_{\mu \nu}$, $K_{\mu \nu}$ and $D_{\mu \nu}$ off-shell, we 
need to re-examine the $\mathcal{Q}$-transformations of these fields.
The on-shell $\mathcal{Q}$-transformations of $\Lambda_{\mu \nu}$,
$\bar{\Lambda}_{\mu \nu}$ are 
\begin{align}
\mathcal{Q} \Lambda_{\mu \nu} = 2 U_{\mu \nu}, \quad 
\mathcal{Q} \bar{\Lambda}_{\mu \nu} = 2 V_{\mu \nu},
\label{eq:calQ_transf}
\end{align}
where we have defined 
\begin{align}
U_{\mu \nu}
&\equiv
-
uF_{\mu\nu}^{+}+
\frac{i}{2} u [\varphi_{\mu},\varphi_{\nu}]^{+}
+\frac{1}{\sqrt{2}} v(D_{\mu}\varphi_{\nu}-D_{\nu}\varphi_{\mu})^{+},
\notag \\
V_{\mu \nu}
&\equiv -
vF_{\mu\nu}^{-}
+\frac{i}{2} v[\varphi_{\mu},\varphi_{\nu}]^{-}
- \frac{1}{\sqrt{2}} u(D_{\mu}\varphi_{\nu}-D_{\nu}\varphi_{\mu})^{-}.
\end{align}
We modify the transformation \eqref{eq:calQ_transf} as
\begin{align}
\mathcal{Q} \Lambda_{\mu \nu} &= 2 \sqrt{u^2+v^2} K_{\mu \nu} + 2 U_{\mu
 \nu}, \notag \\
\mathcal{Q} \bar{\Lambda}_{\mu \nu} &= 2 \sqrt{u^2+v^2} D_{\mu \nu} + 2
 V_{\mu \nu}.
\label{eq:calQ_lambda}
\end{align}
The transformations of $K_{\mu \nu}$ and $D_{\mu \nu}$ are determined as
in the $Q$- and $\bar{Q}$-transformations. 
We redefine the auxiliary fields as 
\begin{align}
\mathcal{G}_{\mu\nu}
&=
\sqrt{u^{2}+v^{2}}\,K_{\mu\nu}+U_{\mu\nu},
\notag \\
\mathcal{H}_{\mu\nu}
&=
\sqrt{u^{2}+v^{2}}\,D_{\mu\nu}+V_{\mu\nu}, 
\end{align}
so that \eqref{eq:calQ_lambda} becomes a simple form as 
\begin{align}
\mathcal{Q} \Lambda_{\mu\nu} =
2\mathcal{G}_{\mu\nu},  \quad 
\mathcal{Q} \bar{\Lambda}_{\mu\nu} =
2\mathcal{H}_{\mu\nu}.
\label{eq:calQ_lambda_redef}
\end{align}
Then the transformations of $\mathcal{G}_{\mu \nu}$ and
$\mathcal{H}_{\mu \nu}$ are 
\begin{align}
\mathcal{Q} \mathcal{G}_{\mu\nu}&=
(u^{2}+v^{2})\Bigl(\sqrt{2}i[\varphi,\Lambda_{\mu\nu}]
-\sqrt{2}\Omega^{\lambda}D_{\lambda}\Lambda_{\mu\nu}
+\sqrt{2}(\Omega_{\mu}{}^{\lambda}\Lambda_{\lambda\nu}
-\Omega_{\nu}{}^{\lambda}\Lambda_{\lambda\mu})
\Bigr),
\notag\\
\mathcal{Q} \mathcal{H}_{\mu\nu}&=
(u^{2}+v^{2})\Bigl(\sqrt{2}i[\varphi,\bar{\Lambda}_{\mu\nu}]
-\sqrt{2}\Omega^{\lambda}D_{\lambda}\bar{\Lambda}_{\mu\nu}
+\sqrt{2}(\Omega_{\mu}{}^{\lambda}\bar{\Lambda}_{\lambda\nu}
-\Omega_{\nu}{}^{\lambda}\bar{\Lambda}_{\lambda\mu})
\Bigr).
\label{eq:Marcus_auxiliary_linear}
\end{align}
The $\mathcal{Q}$-transformations of the other fields are 
obtained from \eqref{eq:calQ}. 

Now we construct the gauge fermion $\widehat{\Xi}$ which satisfies 
$S_3 = \mathcal{Q} \widehat{\Xi}$.
Since we have changed the transformations of $\Lambda_{\mu \nu}$, 
$\bar{\Lambda}_{\mu \nu}$, $K_{\mu \nu}$ and $D_{\mu \nu}$, 
we decompose the gauge fermion as 
$\widehat{\Xi} = \widehat{\Xi}^{(1)} + \widehat{\Xi}^{(2)}$, 
where $\widehat{\Xi}^{(1)}$ is the linear terms in $\Lambda_{\mu\nu}$ 
and $\bar{\Lambda}_{\mu\nu}$ and $\widehat{\Xi}^{(2)}$ does not contain 
these fields.
Using the transformations \eqref{eq:calQ_lambda},
\eqref{eq:Marcus_auxiliary_linear} and the $\mathcal{Q}$-transformations
of the other fields, we find 
\begin{align}
\widehat{\Xi}^{(1)}
&=
\frac{1}{u^2+v^2}
\int\!d^{4}x\,
\frac{1}{\kappa g^{2}}\textrm{Tr}\biggl[
\biggl(\frac{1}{2}U_{\mu\nu}\Lambda^{\mu\nu}
-\frac{1}{2}\mathcal{G}_{\mu\nu}\Lambda^{\mu\nu}\biggr)
+\biggl(\frac{1}{2}V_{\mu\nu}\bar{\Lambda}^{\mu\nu}
-\frac{1}{2}\mathcal{H}_{\mu\nu}\bar{\Lambda}^{\mu\nu}\biggr)
\biggr].
\label{eq:Marcus_Xi1}
\end{align}
In order to find $\widehat{\Xi}^{(2)}$ we take the following ansatz
\begin{gather}
\widehat{\Xi}^{(2)} = a \bar{\Xi}_{3}' + b \Xi_{3}'.
\label{eq:ansatz}
\end{gather}
Here $a$, $b$ are constants and  $\Xi_{3}'$, $\bar{\Xi}_3'$ are terms that do not contain 
$\Lambda_{\mu \nu},\bar{\Lambda}_{\mu \nu}$ in $\Xi_3$ and $\bar{\Xi}_3$ 
respectively.
Using the supersymmetry transformations \eqref{eq:Marcus_OS1}, \eqref{eq:Marcus_OS2}, 
we can show that $\Xi_{3}'$ and $\bar{\Xi}_3'$ are the exact forms as 
\begin{align}
\Xi_{3}' = Q V,\quad \bar{\Xi}'_{3} = - \bar{Q} V, 
\label{eq:Marcus_exact}
\end{align}
where $V$ is given by 
\begin{align}
V=
\int\!d^{4}x\,
\frac{1}{\kappa g^{2}}\textrm{Tr}\biggl[
\frac{1}{8}\Lambda\bar{\Lambda}-\frac{1}{4}\varphi^{\mu}
(D_{\mu}\bar{\varphi}-F_{\mu\nu}\bar{\Omega}^{\nu})
\biggr].
\label{eq:V_define}
\end{align}
We can find the constants $a$, $b$ such that 
\begin{align}
\widehat{\Xi}^{(2)}
=\frac{1}{u^{2}+v^{2}}(-u \bar{Q}+v Q )V,
\end{align}
and the action is written in the $\mathcal{Q}$-exact form. 
We find that the action can be written in the $\mathcal{Q}$-exact form
up to the topological term:
\begin{align}
S_3 = \mathcal{Q}
\biggl(
\widehat{\Xi}^{(1)}
+ \widehat{\Xi}^{(2)}
\biggr)
+\int\!d^{4}x\,
\frac{1}{\kappa g^{2}}\textrm{Tr}\biggl[
\frac{u^{2}-v^{2}}{4(u^{2}+v^{2})}F_{\mu\nu}\tilde{F}^{\mu\nu}
\biggr].
\label{eq:comb_Q-exact}
\end{align}
The dependence on $u$ and $v$ of the topological term is the same as the
undeformed case \cite{Kapustin:2006pk}. 
\vspace{0.4cm}
\\
\\
\underline{\textbullet \ $u^2+v^2=0$ case}
\\
\\
In this case, we can choose $(u,v) = (1,i)$.
The supercharge $\mathcal{Q} = Q + i \bar{Q}$ is 
strictly nilpotent without using the gauge transformation and the
Lorentz rotation.
To see this, we introduce the following linear combinations of the
fields \cite{Marcus:1995mq}:
\begin{align}
\mathcal{V}_{\mu}
&=
A_{\mu}+\frac{i}{\sqrt{2}}\varphi_{\mu},
&
\bar{\mathcal{V}}_{\mu}
&=
A_{\mu}-\frac{i}{\sqrt{2}}\varphi_{\mu},
\notag\\
\mathcal{F}_{\mu\nu}
&=
\partial_{\mu}\mathcal{V}_{\nu}-\partial_{\nu}\mathcal{V}_{\mu}
+i[\mathcal{V}_{\mu}, \mathcal{V}_{\nu}],
&
\bar{\mathcal{F}}_{\mu\nu}
&=
\partial_{\mu}\bar{\mathcal{V}}_{\nu}-\partial_{\nu}\bar{\mathcal{V}}_{\mu}
+i[\bar{\mathcal{V}}_{\mu}, \bar{\mathcal{V}}_{\nu}],
\notag\\
\psi_{\mu}
&=
\Lambda_{\mu}-i\bar{\Lambda}_{\mu},
&
\bar{\psi}_{\mu}
&=
\Lambda_{\mu}+i\bar{\Lambda}_{\mu},
\notag\\
\eta
&=
\Lambda-i\bar{\Lambda},
&
\bar{\eta}
&=
\Lambda+i\bar{\Lambda},
\notag\\
\chi_{\mu\nu}
&=
\Lambda_{\mu\nu}-i\bar{\Lambda}_{\mu\nu},
&
G^{+}
&=
G+[\varphi, \bar{\varphi}],
\notag\\
\mathcal{I}_{\mu\nu}
&=
\mathcal{G}_{\mu\nu}-i\mathcal{H}_{\mu\nu},
&
\phi
&=
\varphi-\frac{i}{\sqrt{2}}\Omega^{\mu}\varphi_{\mu}.
\label{eq400}
\end{align}
We note that $\bar{\chi}_{\mu \nu} = \Lambda_{\mu \nu} + i
\bar{\Lambda}_{\mu \nu} 
$ and 
$\bar{\mathcal{I}}_{\mu \nu} = \mathcal{G}_{\mu \nu} + i
\mathcal{H}_{\mu \nu} 
$ are equal to $\tilde{\chi}_{\mu \nu}$ and $\tilde{\mathcal{I}}_{\mu \nu}$, 
respectively. 
From \eqref{eq:Marcus_OS1}, \eqref{eq:Marcus_OS2}, 
\eqref{eq:calQ_lambda_redef} and \eqref{eq:Marcus_auxiliary_linear}, 
the off-shell $\mathcal{Q}$-transformations of these fields become
\begin{align}
\mathcal{Q} \mathcal{V}_{\mu}
&=
2i\psi_{\mu},
&
\mathcal{Q}\psi_{\mu}
&=
0,
\notag\\
\mathcal{Q} \bar{\mathcal{V}}_{\mu}
&=
0,
&&
\notag\\
\mathcal{Q}\bar{\psi}_{\mu}
&=
-4\sqrt{2}i
(\bar{\mathcal{D}}_{\mu}\phi-\bar{\mathcal{F}}_{\mu\nu}\Omega^{\nu}),
&
\mathcal{Q}\phi
&=
0, 
\notag\\
\mathcal{Q}\bar{\varphi}
&=
\sqrt{2}\eta,
&
\mathcal{Q}\eta
&=
0,
\notag\\
\mathcal{Q}\bar{\eta}
&=
4iG^{+},
&
\mathcal{Q}G^{+}
&=
0,
\notag\\
\mathcal{Q}\chi_{\mu\nu}
&=
2\mathcal{I}_{\mu\nu},
&
\mathcal{Q}\mathcal{I}_{\mu\nu}
&=0,
\label{eq:Marcus_transf}
\end{align}
where we have defined the gauge covariant derivatives with respect to the gauge
fields $\mathcal{V}_{\mu}$, $\bar{\mathcal{V}}_{\mu}$ as follows,
\begin{align}
\mathcal{D}_{\mu} * = \partial_{\mu} * +i[\mathcal{V}_{\mu},\ast],\quad
\bar{\mathcal{D}}_{\mu} * =\partial_{\mu} * +i[ \bar{\mathcal{V}}_{\mu},\ast].
\end{align}

Now we examine the gauge fermion $\widehat{\Xi}$, 
which is decomposed into the sum of $\widehat{\Xi}^{(1)}$ and
$\widehat{\Xi}^{(2)}$. 
Starting from the ansatz \eqref{eq:ansatz}, we have 
 $\widehat{\Xi}^{(2)} = - \frac{i}{2} (Q-i\bar{Q})V$, 
where $V$ is given by \eqref{eq:V_define}. 
However we cannot construct $\widehat{\Xi}^{(1)}$ in a similar way as 
\eqref{eq:Marcus_Xi1} since $\mathcal{Q} \mathcal{I}_{\mu \nu} = 0$. 
So, instead of using the transformations \eqref{eq:Marcus_transf}, 
we change the transformation of
$\chi_{\mu \nu}$ by eliminating $\mathcal{I}_{\mu \nu}$ using its 
equation of motion. We define  
\begin{align}
\mathcal{Q}\chi_{\mu\nu}=-2\bar{\mathcal{F}}_{\mu\nu}.
\end{align}
The new $\mathcal{Q}$-transformations are also nilpotent 
off-shell because $\mathcal{Q}\bar{\mathcal{F}}_{\mu\nu}=0$.
With respect to $\mathcal{Q}$ we can take $\widehat{\Xi}^{(1)}$ as 
\begin{align}
\widehat{\Xi}^{(1)}=\int\!d^{4}x\,
\frac{1}{\kappa g^{2}}\textrm{Tr}\biggl[
-\frac{1}{8}\chi^{\mu\nu}\mathcal{F}_{\mu\nu}
\biggr].
\end{align}
Then the action $S_3$ is written as the sum of the $\mathcal{Q}$-exact term and 
the other part:
\begin{align}
S_3 =&
\mathcal{Q} 
\biggl(\widehat{\Xi}^{(1)} + \widehat{\Xi}^{(2)} \biggr) 
+ S_3', 
\\
S_3' =&  \int\!d^{4}x\,
\frac{1}{\kappa g^{2}}\textrm{Tr}\biggl[
-\frac{i}{4}\tilde{\chi}^{\mu\nu}(\bar{\mathcal{D}}_{\mu}\bar{\psi}_{\nu}
-\bar{\mathcal{D}}_{\nu}\bar{\psi}_{\mu})
+\frac{i}{2\sqrt{2}}\tilde{\chi}^{\mu\nu}[\phi, \chi_{\mu\nu}]
\notag\\[2mm]
&\qquad{}
-\frac{1}{2\sqrt{2}}
\Omega^{\lambda}\tilde{\chi}^{\mu\nu}\bar{\mathcal{D}}_{\lambda}\chi_{\mu\nu}
+\frac{1}{2\sqrt{2}}\tilde{\chi}^{\mu\nu}
(\Omega_{\mu}{}^{\lambda}\chi_{\lambda\nu}
-\Omega_{\nu}{}^{\lambda}\chi_{\lambda\mu})
\biggr].
\end{align}
Here $S_3'$ is not $\mathcal{Q}$-exact but $\mathcal{Q}$-closed.
The deformed terms in $S'_3$ are obtained from the undeformed one 
by using \eqref{eq:shift}.
Although the undeformed part is independent of the metric 
\cite{Marcus:1995mq}, the deformed part depends on the metric through
$\Omega_{\mu \nu}$.

\section{Conclusion and discussions}
In this paper, we have constructed the off-shell scalar supersymmetry
associated with the three different topological twists
 in the $\Omega$-deformed ${\cal N}=4$ super Yang-Mills theory. 
The scalar supercharges form the closed algebra
up to the gauge transformation, the Lorentz rotation associated
with the $\Omega$-vector fields, and the flavor rotation.
We have shown that the $\Omega$-deformed action is written in the exact
form with respect to the scalar supercharges up to topological terms
except the case of the Marcus twist with $u^2+v^2=0$. 
The twisted ${\cal N}=4$ super Yang-Mills theories can be 
naturally deformed in the $\Omega$-background.

It would be important to study the quantum aspects of the deformed theory
since the fixed point equations for the scalar supersymmetry are
deformed by the $\Omega$-background, which could change the partition function. 
For the half twist, the partition function is indeed the same 
as the $\mathcal{N}=2^*$ deformation \cite{Nekrasov:2003rj}.
For the Vafa-Witten twist, this would be a generalization of 
\cite{Bruzzo:2002xf}.
Furthermore it would be an interesting  problem to study the S-duality 
of the $\Omega$-deformed $\mathcal{N}=4$ theory.

In \cite{Ito:2012hs} we showed that the $\Omega$-deformed
$\mathcal{N}=4$ theory has other on-shell supersymmetry associated with
the tensor supercharges. 
It would be interesting to study the off-shell structure of the supersymmetry 
and its realization in 
dimensionally reduced theory \cite{Berkovits:1993hx,Evans:1994np}.
We have 
also studied the deformed supersymmetries in the Nekrasov-Shatashvili limit 
\cite{Nekrasov:2009rc}.
In this limit, on-shell supersymmetry is enhanced to $\mathcal{N} =
(2,2)$ supersymmetry in the case of the half twist, 
$\mathcal{N} = (4,4)$ supersymmetry in the Vafa-Witten
and the Marcus twists, where by the notation 
${\cal N}=(m,n)$ we mean that the theory has $m$ chiral and $n$
anti-chiral supercharges.
It is interesting to study the off-shell transformations of these 
supersymmetries and their BPS states \cite{Hellerman:2012rd}.
In particular, one can study the BPS equations 
 in the Nekrasov-Shatashivili limit, which has been
investigated in the $\mathcal{N}=2$ case \cite{Ito:2011ta}.

\subsection*{Acknowledgements} 
The work of K.~I.  is supported 
in part by Grant-in-Aid for Scientific Research from the Japan Ministry of Education, 
Culture, Sports, Science and Technology. 
The work of H.~N. is supported in part by National Science Council
and National Center for Theoretical Sciences, Taiwan, R.O.C. 
The work of S.~S is supported in part by Sasakawa Scientific Research
Grant from The Japan Science Society and Kitasato University Research
Grant for Young Researchers.

\begin{appendix}
 \section{Dirac matrices in four and six dimensions}
The four-dimensional sigma matrices 
$\sigma^{\mu}$, $\bar{\sigma}^{\mu}$ are defined by 
$\sigma^{\mu} = (i \tau_1, i \tau_2, i \tau_3, \mathbf{1}_2)$,
$\bar{\sigma}^{\mu} = (- i \tau_1, - i \tau_2, - i \tau_3,
\mathbf{1}_2)$ where $\tau_{\tilde{c}} \ (\tilde{c} = 1,2,3)$ are the
Pauli matrices. The four-dimensional Lorentz generators are defined by 
$\sigma^{\mu \nu} = \frac{1}{4} (\sigma^{\mu} \bar{\sigma}^{\nu} -
\sigma^{\nu} \bar{\sigma}^{\mu})$, $\bar{\sigma}^{\mu \nu} = \frac{1}{4}
(\bar{\sigma}^{\mu} \sigma^{\nu} - \bar{\sigma}^{\nu} \sigma^{\mu})$.

The Dirac matrices $(\Sigma_{a})^{AB}$ and $(\bar{\Sigma}_{a})_{AB}$ in six dimensions are defined by 
\begin{align}
\Sigma_{1}
&=
\begin{pmatrix} i\tau^{2} & 0 \\ 0 & i\tau^{2} \end{pmatrix},
&
\Sigma_{2}
&=
\begin{pmatrix} \tau^{2} & 0 \\ 0 & -\tau^{2} \end{pmatrix},
&
\Sigma_{3}
&=
\begin{pmatrix} 0 & -\tau^{3} \\ \tau^{3} & 0 \end{pmatrix},
\notag\\[2mm]
\Sigma_{4}
&=
\begin{pmatrix} 0 & i\boldsymbol{1}_{2} \\ -i\boldsymbol{1}_{2} & 0 
\end{pmatrix},
&
\Sigma_{5}
&=
\begin{pmatrix} 0 & - \tau^{1} \\ \tau^{1} & 0 \end{pmatrix},
&
\Sigma_{6}
&=
\begin{pmatrix} 0 & \tau^{2} \\ \tau^{2} & 0 \end{pmatrix},
\notag\\[2mm]
\bar{\Sigma}_{1}
&=
\begin{pmatrix} -i\tau^{2} & 0 \\ 0 & -i\tau^{2} \end{pmatrix},
&
\bar{\Sigma}_{2}
&=
\begin{pmatrix} \tau^{2} & 0 \\ 0 & -\tau^{2} \end{pmatrix},
&
\bar{\Sigma}_{3}
&=
\begin{pmatrix} 0 & \tau^{3} \\ -\tau^{3} & 0 \end{pmatrix},
\notag\\[2mm]
\bar{\Sigma}_{4}
&=
\begin{pmatrix} 0 & i\boldsymbol{1}_{2} \\ -i\boldsymbol{1}_{2} & 0 
\end{pmatrix},
&
\bar{\Sigma}_{5}
&=
\begin{pmatrix} 0 &  \tau^{1} \\ - \tau^{1} & 0 \end{pmatrix},
&
\bar{\Sigma}_{6}
&=
\begin{pmatrix} 0 & \tau^{2} \\ \tau^{2} & 0 \end{pmatrix}.
\end{align}

\end{appendix}

\end{document}